\documentclass[preprint]{aastex}
\usepackage{emulateapj5,apjfonts,epsfig}

\newcommand{\mr}{MR~2251--178}

\begin{document}

\title{A Search for Very Extended Ionized Gas in Nearby Starburst and Active Galaxies}

\author{S. Veilleux}
\affil{Department of Astronomy, University of Maryland, College Park,
MD 20742; \\ veilleux@astro.umd.edu}

\author{P. L. Shopbell}
\affil{Department of Astronomy, California Institute of Technology, 
Pasadena, CA, 91125}

\author{D. S. Rupke\altaffilmark{3}}
\affil{Department of Astronomy, University of Maryland, College Park,
MD 20742}

\author{J. Bland-Hawthorn\altaffilmark{5}}
\affil{Anglo-Australian Observatory, P.O. Box 296, Epping, NSW 2121,
Australia}

\author{G. Cecil\altaffilmark{3}}
\affil{Department of Physics and Astronomy, CB \#3255, 
University of North Carolina, Chapel Hill, NC 27599-3255}

\altaffiltext{1}{Current Address: 320-47 Downs Lab., Caltech, Pasadena,
CA 91125 and Observatories of the Carnegie Institution of Washington,
813 Santa Barbara Street, Pasadena, CA 91101; veilleux@ulirg.caltech.edu}

\altaffiltext{2}{Cottrell Scholar of the Research Corporation}

\altaffiltext{3}{Visiting Astronomer, Anglo-Australian Observatory,
P.O.  Box 296, Epping, NSW 1710, Australia}

\altaffiltext{4}{Visiting Astronomer, Mount Stromlo and Siding Springs
Observatory, operated by the Research School of Astronomy and
Astrophysics, Cotter Road, Weston Creek, Canberra, ACT72611, Australia}

\altaffiltext{5}{Visiting Astronomer, William Herschel Telescope, 
Isaac Newton Group of Telescopes, Observatorio Roque de Los Muchachos, 
Santa Cruz de la Palma, Canary Islands E-38700, Spain}

\begin{abstract}
We report the results from a pilot study of 10 nearby starburst and
active galaxies conducted with the Taurus Tunable Filter (TTF) on the
Anglo-Australian and William Herschel Telescopes. The main purpose of
this imaging survey is to search for warm emission-line gas on the
outskirts (galactocentric distances $R$ $\ga$ 10 kpc) of galaxies to
provide direct constraints on the size and geometry of the ``zone of
influence'' of these galaxies on their environment. Gaseous complexes
or filaments larger than $\sim$ 20 kpc are discovered or confirmed in
six of the galaxies in the sample (NGC~1068, NGC~1482, NGC~4388,
NGC~6240, NGC~7213, and \mr). Slightly smaller structures are seen for
the first time in the ionization cones and galactic winds of NGC~1365,
NGC~1705, Circinus galaxy, and ESO484-G036. The TTF data are combined
with new optical long-slit spectra as well as published and archived
radio and X-ray maps to constrain the origin and source of ionization
of these filaments. A broad range of phenomena is observed, including
large-scale ionization cones and galactic winds, tidal interaction,
and ram-pressure stripping by an intracluster medium. The source of
ionization in this gas ranges from shock ionization to photoionization
by the central AGN or in-situ hot young stars. The sample is too small
to draw statistically meaningful conclusions about the extent and
properties of the warm ionized medium on large scale and its relevance
to galaxy formation and evolution. The next generation of tunable
filters on large telescopes promises to improve the sensitivity to
faint emission-line fluxes at least tenfold and allow systematic
surveys of a large sample of emission-line galaxies.
\end{abstract}


\section{Introduction}

The need for a comprehensive survey of the warm ionized medium in the
local universe can hardly be overstated. This gas phase may contribute
significantly to the local baryon budget (e.g., Fukugita, Hogan, \&
Peebles 1998), but very little is known about its distribution.  While
an important fraction of this material is almost certainly in the form
of intergalactic clouds not related to any individual galaxy, some may
inhabit the dark matter halos of galaxies (e.g., Rauch 1998).

This gas phase is a key witness to galaxy formation and evolution. In
a hierarchical CDM universe (e.g., Jenkins et al.  1998) most of the
activity associated with galaxy formation takes place at $z$ $\ga$ 1,
except perhaps in the outer reaches of galaxies where ``primordial''
gas may still be accreting today. Some of this material will
necessarily be ionized by the metagalactic ionizing radiation, and
possibly also by local sources of ionizing radiation such as active
galactic nuclei (AGN) and starbursting stellar populations. In this
picture, the warm ionized gas on the outskirts of galaxies represents
left-over debris associated with galaxy formation (e.g., Bland-Hawthorn
1999; Blitz et al. 1999).

The warm ionized gas is also an excellent probe of the feedback
processes taking place in galaxies. Ionizing radiation and mechanical
energy from star-forming regions and quasars may severely limit the
amount of star formation and affect galaxy evolution. The emerging
radiation field defines regions where gas is warm and ionized (e.g.,
``proximity effect'' in quasars and Lyman-break galaxies and
``ionization cones'' in nearby active and starburst galaxies; e.g.,
Bajtlik, Duncan, \& Ostriker 1988; Steidel, Pettini, \& Adelberger
2001; Wilson \& Tsvetanov 1994). Galactic winds may blow out through
the gaseous halos of galaxies and into the IGM, enriching and heating
the intergalactic environment in the process. These winds may be
responsible for the well-known mass-metallicity relation in galaxies
(e.g., Larson \& Dinerstein 1975; Vader 1986; Franx \& Illingworth
1990). The discovery of H$\alpha$ filaments and diffuse soft X-ray
emission out to 11.6 kpc from the prototypical starburst/superwind
galaxy M82 (Devine \& Bally 1999; Lehnert, Heckman, \& Weaver 1999;
Stevens, Read, \& Bravo-Guerrero 2003) emphasizes the need for
surveying large areas around superwind galaxies to constrain the size,
energetics, and impact of these superwinds. The large ``cavities'' in
the X-ray surface brightness of several cooling flow clusters with
radio-loud cD galaxies (e.g., B\"ohringer et al. 1993; Fabian et
al. 2000; McNamara et al. 2000; Fabian 2001; Quilis, Bower, \& Balogh
2001; Heinz et al. 2002) point to equally large zones of influence for
AGN-driven outflows. Galactic winds in the early universe are likely
to have had an even stronger influence on the environment (e.g.,
Alderberger et al. 2003; Shapley et al. 2003).

The present paper describes the results from a pilot survey of 10
nearby starburst and active galaxies with the Taurus Tunable Filter
(TTF) on the Anglo-Australian and William Herschel Telescopes. The
main goals of this survey are to search for warm (T $\approx$ 10$^4$
K) line-emitting gas on the outskirts of these galaxies, and study the
properties of this gas to constrain its origin and overall importance.
The names, redshifts, and nuclear types of the galaxies in the sample
are listed in Table 1 along with a summary of the findings from the
TTF survey.  These galaxies were selected based on the fact that they
all present ionization cones and/or galactic winds on $\sim$ kpc
scale. This method of selection and the small number of galaxies in
the sample imply that this sample is probably not representative of
the local population of starburst/active galaxies, and therefore
should not be used for statistical purposes.  Nevertheless, the
results from this pilot survey should provide important clues on the
range of phenomena taking place in the general local galaxy
population.

This paper is organized as follows. In \S 2, we discuss the methods of
observations.  The results are described individually for each object
in \S 3. In \S 4, we summarize the results and discuss future avenues
of research. The present discussion complements recent Fabry-Perot
searches for warm ionized gas on the outskirts of ``normal'' galaxies
including our own Galaxy (e.g., Bland-Hawthorn et al. 1998; Putman et
al. 2003) and several dwarf, spiral, and elliptical galaxies (e.g.,
Bland-Hawthorn, Freeman, \& Quinn 1997a; Meurer et al. 1999; Ferguson,
van der Hulst, \& van Gorkom 2001; Miller \& Veilleux 2003a, 2003b),
as well as in isolated extragalactic H~I clouds (e.g., Weymann et
al. 2001) and a few active galaxies with collimated jet outflows
(e.g., Cecil et al. 2000; Tadhunter et al. 2000). The results in the
present paper supersede those from Veilleux (2002). The results for
two of the galaxies in the sample have already been presented
elsewhere (NGC~1482: Veilleux \& Rupke 2002; \mr: Shopbell, Veilleux,
\& Bland-Hawthorn 1999), but are discussed again in the present paper
for the sake of completeness and to add to the discussion (e.g., a new
unpublished X-ray map of NGC~1482 are presented here along with a
deeper reprocessed H$\alpha$ + [N~II] $\lambda$6583 image of this
object).

\section{Observations}

The TTF on the 3.9-meter Anglo-Australian Telescope (AAT) was used for
all but one of the observations reported in the present paper. The TTF
on the 4.2-meter William Herschel Telescope (WHT) was used for
NGC~1068. The TTF is described in detail in Bland-Hawthorn \& Jones
(1998) and Bland-Hawthorn \& Kedziora-Chudczer (2003). The data
acquisition and reduction techniques used to reach low flux levels
with the TTF have already been discussed in several papers (e.g.,
Shopbell et al. 1999; Glazebrook \& Bland-Hawthorn 2001; Veilleux \&
Rupke 2002; Jones, Shopbell, \& Bland-Hawthorn 2002; Miller \&
Veilleux 2003a). The TTF is uniquely suited to carry out deep searches
for emission-line gas on the outskirts of galaxies, combining wide
field of view ($\sim$ 10$\arcmin$) with outstanding narrow-band
imaging capabilities over a broad range in wavelength (3500 \AA\ --
1.0 $\mu$m) and bandpass (10 -- 100 \AA). A summary of the TTF
observations is given in Tables 2 and 3.  Nearly all of the AAT
observations were carried out in the ``charge shuffling/frequency
switching'' mode, where the charges are moved up and down within the
detector at the same time as switching between two discrete
frequencies with the tunable filter.  The charges were generally moved
every minute and the chip was read after typically spending 32 minutes
of integration time (16 minutes on-band and 16 minutes off-band). As a
result, the continuum and emission-line images are produced nearly
simultaneously, therefore averaging out temporal variations associated
with atmospheric lines and transparency, seeing, instrument and
detector instabilities.  Most of the AAT images were obtained in a
straddle mode, where the off-band image is made up of a pair of images
that ``straddle'' the on-band image in wavelength (e.g., $\lambda_1$ =
6500 \AA\ and $\lambda_2$ = 6625 \AA\ for rest-frame H$\alpha$); this
greatly improves the accuracy of the continuum removal since it
corrects for slopes in the continuum and underlying absorption
features. The charge shuffling and frequency switching capabilities
were not available at the WHT. The flux levels reached by the TTF
observations are listed in the last column of Table 3.

Long-slit optical spectra were also obtained for some of the objects
to clarify the origin and source of ionization of the warm ionized
material. Tables 2 and 4 summarize the details of these observations.
These spectra were reduced using standard IRAF routines.  When
available, complementary X-ray and radio maps were used to track the
hot (X-rays), relativistic (20-cm) and neutral (HI) gas phases in the
sample galaxies and allow us to draw a more complete picture of these
objects.

\section{Results}

\subsection{NGC~1068}

NGC~1068 is arguably the best studied active galaxy in the local
universe.  It is a Sb galaxy which is nearly face-on ($i \approx 30$;
de Vaucouleurs et al. 1991) and shows evidence for a $\sim$ 3-kpc bar
(e.g., Scoville et al. 1988; Thronson et al. 1989; Helfer \& Blitz
1995). Signs of the AGN in the core of this galaxy are evident at
nearly all wavelengths.  It was one of the six objects originally
studied by Seyfert (1943) with bright optical ``emission lines similar
to those in planetary nebulae.''  Four decades later,
spectropolarimetry of NGC~1068 revealed the presence of broad (FWHM
$\approx$ 4,500 km s$^{-1}$) recombination lines in scattered light,
suggesting for the first time that the $\la$ pc-scale broad line
region in this object is obscured by an optically thick torus (e.g.,
Miller \& Antonucci 1983; Antonucci \& Miller 1985; Miller, Goodrich,
\& Mathews 1991). Indirect evidence for a inner disk structure in
NGC~1068 comes from the presence of bright cones of photoionized gas
detected at optical wavelengths (e.g., Baldwin, Wilson, \& Whittle
1987; Pogge 1988a; Macchetto et al. 1994; Kraemer, Ruiz, \& Crenshaw
1998) and also in X-rays (e.g., Young, Wilson, \& Shopbell 2001;
Kinkhabwala et al. 2002).  The ionization cones are roughly aligned
with the inner radio jet and large-scale radio structure (e.g., Wilson
\& Ulvestad 1987; Gallimore et al. 1996). Most of the inner
line-emitting clouds originally discovered by Walker (1968) are taking
part in a large-scale outflow event (e.g., Cecil, Bland, \& Tully
1990; Arribas, Mediavilla, \& Garcia-Lorenzo 1996; Crenshaw \& Kraemer
2000). Some of the line-emitting material near the nucleus is
outflowing at velocities of up to $\sim$ 3200 km s$^{-1}$ (Cecil et
al. 2002b); this material appears to be accelerated radiatively by the
AGN (Dopita et al. 2002). UV line ratios are inconsistent with shock
excitation (Groves et al., in prep.). 

The anisotropic radiation field from the AGN in NGC~1068 is already
known to affect the inner $\sim$ 10 kpc diameter disk of the host
galaxy (Bland-Hawthorn, Sokolowski, \& Cecil 1991; Sokolowski,
Bland-Hawthorn, \& Cecil 1991). Our new data on NGC~1068 now show that
the ionization cone extends even further.  Figure 1 presents deep
H$\alpha$ and [O~III] $\lambda$5007 images obtained with the TTF. A
complex of knots and filaments are detected out to $R \approx$ 11 kpc
along P.A. $\approx$ 15 -- 70$^\circ$. Another fainter complex is seen
out to $\sim$ 10 kpc in the opposite direction at P.A. $\approx$ 200
-- 260$^\circ$. There is a deficit of filaments in all other
directions.

Figure 2 compares the locations of the filaments with an unpublished
H~I 21-cm map graciously provided to us by E. Brinks (2003; private
communication).  Some of the optical filaments (especially those to
the south-west) lie slightly beyond the sharp H~I ``edge'' of
NGC~1068. As in the case of NGC~253 (Bland-Hawthorn et al. 1997a), the
filaments are too bright to be photoionized solely by the metagalactic
ionizing radiation -- another source of ionization is needed.  The
surprisingly bright H$\alpha$ emission beyond the H~I edge of NGC~253
is probably due to photoionization by hot young stars in the inner
disk (Bland-Hawthorn et al. 1997a). We suspect a different origin for
the filaments in NGC~1068. As is clearly apparent in Figure 1, these
filaments are contained within a biconical region which is roughly
aligned with the nuclear outflow/jets and the optical ionization cone
seen on smaller scales.  Figure 2 also shows that there is a good
(though not perfect) match in P.A. between the X-ray emission and the
optical filaments on large scale. This biconical geometry on small and
large scales strongly suggests that the AGN is responsible for the
ionization of the gas on both scales.

The large [O~III]/H$\alpha$ ratios ($\ga$ 1; Fig. 1) observed in the
gas within the large-scale cones seem to support this scenario (i.e.
the hard radiation field from the AGN contributes to the high
ionization level of the gas in this region).  Deep MSSSO and AAT
long-slit spectra of the north-east filament at $R \approx$ 6 -- 11
kpc confirm the role of the AGN. The line ratios measured along this
filament are plotted in Figure 3.  Using the diagnostic tools of
Veilleux \& Osterbrock (1987), we find that most of the line ratios
are Seyfert-like. No obvious systematic gradient in the line ratios is
detected along the brighter portion of the filament.  The $\sim$ 100
km s$^{-1}$ blueshift and slight velocity gradient apparent in the
data (particularly in the higher S/N data obtained at MSSSO; see
Fig. 4) are consistent with gas in rotation in the P.A. $\approx$
79$^\circ$ galactic disk (e.g., Galletta \& Recillas-Cruz 1982; Kaneko
et al. 1992; Sofue 1997; Sofue et al. 1999). The narrow line widths in
the brighter portion of the filament (FWHM $<$ 200 km s$^{-1}$)
suggest that quiescent gas from the disk is being photoionized by the
central AGN.  The apparent broadening in the emission-line profiles
extracted from the fainter (southern) portion of the filament (FWHMs
reach $\sim$ 300 -- 400 km s$^{-1}$ in some locations and maybe
accompanied by line splitting; see Fig. 4) may indicate that shocks
become important or the AGN is photoionizing a spray of gas with a
broader range of line-of-sight velocities.  The elevated
[N~II]/H$\alpha$ and [S~II]/H$\alpha$ ratios in this region (not shown
in Fig. 3 because [O~III]/H$\beta$ is poorly constrained) could be due
to either shocks or AGN photoionization with low ionization parameter
(= ratio of the density of ionizing photons to that of electrons).
Intervening disk material may explain the slightly fainter emission in
the south-west cone since the radiation field emerges above the disk
in the north-east quadrant and below the disk in the south-west
quadrant (e.g., Cecil et al. 1990; Bland-Hawthorn et al. 1997b).

\subsection{NGC~1365}

NGC~1365 is a giant barred Sb galaxy in the Fornax cluster, host to a
beautiful grand design spiral structure and a Seyfert 1.5 nucleus.  It
has been the subject of several spectroscopic studies, including the
pioneering work of Burbidge \& Burbidge (1960) who were the first to
point out the possibility of a large-scale outflow in this object.
Line splitting of the [O~III] profiles observed by Phillips et
al. (1983) and J\"ors\"ater, Lindblad, \& Boksenberg (1984) confirmed the
complex velocity field of the gas in the nuclear region, and revealed
the highly ionized state of the outflowing material. Subsequent
studies have traced the extent and geometry of the highly ionized
outflowing material (e.g., Edmunds et al. 1988; Storchi-Bergmann \&
Bonatto 1991; Hjelm \& Lindblad 1996; Kristen et al. 1997). An
accelerated outflow in a hollow biconical structure with opening angle
$\sim$ 100$^\circ$ has been suggested (Hjelm \& Lindblad 1996). This
structure is roughly aligned with a jet-like radio feature observed by
Sandqvist, J\"ors\"ater, \& Lindblad (1995) and Morganti et al. (1999).

TTF H$\alpha$, [N~II] $\lambda$6583, and [OIII] $\lambda$5007 images
were obtained of NGC~1365; they are presented in Figures 5 -- 8. The
[O~III] data confirm the presence of the bright south-east plume of
[O~III]-emitting material seen in previous studies, but also show for
the first time fainter plumes of material on the opposite side of the
nucleus. The fainter [O~III] emission in the TTF data extends over
$\sim$ 1$\arcmin$ , and is much less conical than the bright SE
plume. The [O~III] $\lambda$5007/H$\alpha$ ratios in this region are
larger than unity, suggesting photoionization by the AGN. Strings of
bright H~II regions and complex dust lanes bisect the [O~III]
structure and lowers the [O~III]/H$\alpha$ ratios along a diagonal
line passing immediately north of the nucleus.

The [N~II]/H$\alpha$ line ratio maps shown in Figures 6 and 8 indicate
that the zone of influence of the AGN in NGC~1365 extends further than
suspected.  [N~II] $\lambda$6583/H$\alpha$ ratios in excess of unity
are seen out to $\sim$ 1$\arcmin$ ($\sim$ 5 -- 6 kpc) from the
nucleus. Surprisingly, the high-[N~II]/H$\alpha$ region appears to be
almost rectangular rather than biconical, although the axis of
symmetry is the same as that of the high [O~III]/H$\alpha$ structure
(P.A. $\approx$ --45$^\circ$). It is not clear at present whether this
fainter material is also taking place in the bright outflow studied by
Hjelm \& Lindblad (1996), or is simply ambient disk material being
illuminated by the central AGN.
 
\subsection{NGC~1482}

The results from our TTF imaging and optical long-slit spectroscopy of
NGC~1482 have already been published in Veilleux \& Rupke (2002).
These data reveal a starburst-driven wind extending 1.5 kpc above and
below the disk of the host galaxy with expansion velocities on the
order of $\sim$ 250 km s$^{-1}$ and kinetic energy of at least 2
$\times$ 10$^{53}$ ergs. The TTF data are presented again in Figure
9. The [N~II] $\lambda$6583/H$\alpha$ ratio map derived from these
data shows that the entrained wind material has [N~II]
$\lambda$6583/H$\alpha$ ratios in excess of unity while the disk
material is characterized by H~II region-like line ratios indicative
of the starburst. Shock ionization is suspected to be responsible for
the LINER-like line ratios of the entrained material.

The new X-ray map presented in Figure 9 brings credence to this
scenario.  This map is based on the unpublished archived 28.56 ks
exposure (P.I. D. Strickland) with the ACIS-S3 detector on the {\em
Chandra X-ray Observatory}.  High-background data were filtered out of
the original image; the effective exposure time is 24.40 ks. A fixed
Gaussian kernel with $\sigma$ = 3 pixels (1$\farcs$5) was used for
smoothing.  The energy range of the data is 0.2 -- 3 keV.  The X-ray
emission (especially at low energies) is elongated along the minor
axis of the galaxy and shows a remarkably good match with some of the
extraplanar optical filaments. A tight spatial correlation between
X-ray emitting material and optical emission-line gas is also observed
in two other well-known galactic winds studied in detail with {\em
Chandra}: NGC~253 (Strickland et al. 2000) and NGC~3079 (Cecil et
al. 2001; Cecil, Bland-Hawthorn, \& Veilleux 2002a). In both of these
cases, the data suggest that the superwind has driven cool disk gas
into the halo, with X-rays being emitted either as upstream, standoff
bow shocks or by cooling at cloud/wind conductive interfaces. The same
phenomenon appears to be taking place in NGC~1482 (and NGC~6240; see
\S 3.6).  Good correlations between extraplanar H$\alpha$ and X-ray
emission have been known to exist in a number of star-forming galaxies
(e.g., Dahlem et al. 2003 and references therein), but the new
$Chandra$ data now show that this correlation extends down to $\sim$
1\arcsec\ scale.

Reprocessing of the TTF data now reveals that the galactic wind in
this galaxy extends much further than first suspected by Veilleux \&
Rupke (2002; see Fig. 10). The summed H$\alpha$ + [N~II] $\lambda$6583
image now clearly shows the presence of a filament that extends out to
$\sim$ 7 kpc north-west of the nucleus.  Another filament extending
out to $\sim$ 11 kpc north-east of the nucleus is visible in Figure
10, although scattered light from a bright star in the field makes the
detection of this filament uncertain. Diffuse emission near the
detection limit of our data also appears to be present $\sim$ 10 kpc
south of the nucleus. A high-velocity ($\sim$ +400 km s$^{-1}$)
emission-line knot located $\sim$ 3 kpc north of the nucleus is also
visible in [N~II] $\lambda$6583 in the AAT long-slit spectrum
(Fig. 11; this knot is also present at H$\alpha$ but is near the
detection limit of the data).  The discovery of these knots and
filaments will necessarily increase the energetics involved in the
outflow, although the kinematics of the gas on large scale are not yet
fully constrained.

\subsection{NGC~1705}

NGC~1705 is a blue compact dwarf (BCD) galaxy located at 6.2 Mpc
(Meurer et al. 1995). It is host to a starburst-driven wind which is
responsible for the multiple emission-line loops and arcs seen out to
the Holmberg radius, $\sim$ 2.1 kpc (Meurer et al 1992; Hunter,
Hawley, \& Gallagher 1993). Line splitting indicative of expansion
velocities of order $\sim$ 100 km s$^{-1}$ is seen over most of the
line-emitting gas (e.g., Meurer et al.  1992; Marlowe et
al. 1995). The detection of blueshifted UV absorption lines has
confirmed the presence of the wind in this object (Heckman \&
Leitherer 1997; Sahu \& Blades 1997).  Recent $FUSE$ far-UV
spectroscopy has revealed outflowing coronal-phase gas at a few
$\times$ 10$^5$ K, possibly created at the interface of the warm
($\sim$ $10^4$ K) optical line-emitting material and the hot
outrushing gas in a blownout superbubble (Heckman et al. 2001).

The line-emitting loops and arcs seen by previous authors are easily
detected with the TTF in both H$\alpha$ and [N~II] $\lambda$6583
(Fig. 12). Surprisingly, the [N~II]/H$\alpha$ ratio map shows very
little structure across these structures. The [N~II]
$\lambda$6583/H$\alpha$ line ratio peaks at a value of $\sim$ 0.3
near the location of the bright central star cluster (indicated by an
arrow in Fig. 12; this star cluster is NGC~1705-1 in the nomenclature
of Melnick, Moles, \& Terlevich 1985).  [N~II]/H$\alpha$ generally
stays below $\sim$ 0.1 across the remainder of the line-emitting
nebula, except perhaps in a few features where the S/N is low and the
line ratio less reliable.

This line ratio map is very different from that of NGC~1482, even
though both objects host similar starburst-driven winds. As
discussed in the previous section, the large [N~II]/H$\alpha$ ratios
detected in the outflowing gas of NGC~1482 are almost certainly due to
shock excitation. The fact that we do not detect significant
enhancement in the [N~II]/H$\alpha$ map of NGC~1705 may be due to the
fact that the expansion velocities in this dwarf galaxy are considerably
smaller than in NGC~1482, so that shock excitation ($\propto v_{\rm
shock}^3$) is less important than photoionization by the starburst
itself.  This result emphasizes an important limitation of the
excitation technique suggested by Veilleux \& Rupke (2002) to search
for galaxies with starburst-driven winds: the results from this search
technique will necessarily be biased toward powerful, shock-excited
wind structures which show a sharp contrast in excitation properties
relative to the star-forming hosts. Winds in dwarf galaxies will be
harder to detect with this technique. 

\subsection{NGC~4388}

NGC~4388 is a well-known Seyfert 2 galaxy located near the center of
the Virgo cluster (e.g., Phillips \& Malin 1982). The AGN is
responsible for driving a loosely collimated outflow out of the disk
of this edge-on galaxy, visible in the radio (Stone, Wilson, \& Ward
1988; Hummel \& Saikia 1991; Kukula et al. 1995; Falcke, Wilson, \&
Simpson 1998; Irwin, English, \& Sorathia 1999) and optical (e.g.,
Veilleux et al. 1999 and references therein). A recent $Chandra$ study
of NGC~4388 by Iwasawa et al. (2003) now shows direct support for the
galactic outflow in the X-rays. Their map is reproduced in Figure 13
along with the H$\alpha$ and [N~II] $\lambda$6583 images obtained with
the TTF.  Note in Table 3 that the TTF was tuned to a wavelength of
6605 \AA\ near the nucleus of the galaxy -- 13 \AA\ below the
wavelength of H$\alpha$ at systemic velocity -- to map the blueshifted
gas complex north of NGC~4388.  Part of the soft X-ray emission is
clearly produced in the disk, while the extraplanar emission is
directly associated with the outflowing line-emitting material. This
material is highly ionized and therefore presents only modest
[N~II]/H$\alpha$ ratios ($\la$ 0.7; Pogge 1988b; Colina 1992;
Petitjean \& Durret 1993).

A recent study by Yoshida et al. (2002) has revealed emission-line
filaments extending out to $\sim$ 35 kpc from the center of
NGC~4388. Although the origin of these filaments is still uncertain,
stripping of the interstellar medium of NGC~4388 by the ram pressure
of the Virgo intracluster medium is likely to be responsible for some
of these features (see also Veilleux et al. 1999). The recent
discoveries of an isolated H~II region and extended H~I gas in the
vicinity of NGC~4388 by Gerhard et al. (2002) and Vollmer \&
Huchtmeier (2003) bring additional support to the ram-pressure
stripping scenario. The wide field of view of the TTF allows us to
confirm the detection of the filaments (Fig. 14) and provide
additional information on the excitation and kinematics of the
line-emitting material. The [N~II]/H$\alpha$ ratios in the brighter
line-emitting knots beyond a galactocentric distance of $\sim$ 7 kpc
stay below $\sim$ 0.7 but show no obvious monotonic gradient with
distance (Fig. 15). These line ratios do not allow us to identify
unambiguously the source of ionization of the gas (low-velocity
shocks, AGN with low ionization parameter, or in-situ hot stars), but
indicate that the ionization and excitation properties of the brighter
line-emitting knots at these large distances are affected by complex
variations in the ionization parameter or shock velocity rather than
only the distance from the nucleus.  The fact that nearly all of the
emission features seen in the data of Yoshida et al. (2002) are also
detected in our (slightly blueshifted) TTF images suggests that most
of these features have velocities $\la$ +100 km s$^{-1}$ near the
nucleus of the galaxy and $\la$ +250 km s$^{-1}$ in the outer
filaments.

\subsection{NGC~6240}

NGC~6240 is often considered the archetype of luminous infrared
galaxies (log[L$_{\rm IR}$/L$_\odot$] = 11.8); it has been studied
thoroughly at all wavelengths. NGC~6240 is an early merger with two
nuclei separated by $\sim$ 1$\farcs$8 or $\sim$ 1.28 kpc (e.g.,
Carral, Turner, \& Ho 1990; Beswick et al. 2001; Fried \& Schultz
1983; Keel 1990; Bland-Hawthorn, Wilson, \& Tully 1991; Tacconi et
al. 1999; Scoville et al. 2000; Tezca et al. 2000). Unambiguous signs
of an AGN in this system have been detected in the hard X-rays
(Iwasawa \& Comastri 1998; Vignati et al. 1999).  Recent
high-resolution images obtained with $Chandra$ show that both nuclei
are X-ray emitters (Lira et al. 2002) and that both appear to be AGN
(Komossa et al. 2003).  Surrounding the nuclei is a large-scale soft
X-ray nebula possibly powered by a superwind (Schulz et al. 1998;
Komossa, Schulz, \& Greiner 1998; Kolaczyk \& Dixon 2000).  Signs for
a superwind are also seen at optical wavelengths in the form of a
complex of line-emitting filaments and arcs extending over 50 $\times$
60 kpc (Heckman, Armus, \& Miley 1987) and often showing violent gas
motion and LINER-like line ratios (e.g., Heckman et al. 1987; Heckman,
Armus, \& Miley 1990; Keel 1990).

Deep H$\alpha$ and [N~II] $\lambda$6583 images of this object were
obtained with the TTF (Figs. 16 and 17).  H$\alpha$ emission is
detected over an area $\sim$ 70 $\times$ 80 kpc centered on the
nuclei. The distribution of the [N~II] emission differs considerably
from that of H$\alpha$. The incomplete loop and filament located at
P.A. $\approx$ 265 -- 275$^\circ$ and $R \approx 15 - 25$ kpc are
clearly detected in the [N~II] image but are hardly visible in
H$\alpha$; these features have [N~II] $\lambda$6583/H$\alpha$ $\ga$
1.4 and stand out in the excitation map shown in Figure 16.
This large-scale, high-[N~II]/H$\alpha$ structure is roughly aligned
with a 5-kpc X-ray loop detected by Komossa et al. (2003). 
Figure 17 also shows that the brighter portion of this loop coincides
with a emission-line knot with strongly enhanced ($>$ 1) LINER-like
[N~II]/H$\alpha$ line ratios. These results bring further support to
the galactic outflow scenario. The line-emitting gas is entrained in
the starburst-driven wind. Shocks associated with the interaction
between the fast wind and the slow-moving ambient gas contribute to
the heating and ionization of the line-emitting material. The good
match between optical line emission and X-ray emission suggests once
again (see \S 3.3 and references therein) that the X-rays from this
galactic wind are produced either by cooling at the conductive
interfaces between the line-emitting clouds and the wind or as
upstream, standoff bow shocks.

\subsection{NGC~7213}

NGC 7213 is a nearby (22.0 Mpc; $H_0$ = 75 km s$^{-1}$ Mpc$^{-1}$;
Tully 1988) face-on Sa galaxy which is host to a bright Seyfert 1
nucleus with broad H$\alpha$ and strong hard X-ray emission (e.g.,
Phillips 1979; Filippenko \& Halpern 1984). The nuclear spectrum also
exhibits narrow LINER-like emission lines produced by the diffuse gas
around the Seyfert nucleus. Figure 18 shows deep H$\alpha$ and [N~II]
$\lambda$6583 images of this object obtained with the TTF. The TTF
data recover the ring of H~II regions known to exist around the bright
active nucleus of this galaxy (Storchi-Bergmann et al. 1996).  The TTF
image also reveals the presence of a line-emitting filament located
$\sim$ 19 kpc from the nucleus, well outside the optical radius of
this galaxy. This filament was independently discovered by Hameed et
al. (2001), who argues that it is the ionized portion of tidal debris
from a recent merger.

The TTF data suggest that the [N~II]/H$\alpha$ ratios in the filament
are unlike those typically seen in H~II regions (e.g., ring of H~II
regions near the nucleus). The line ratios measured from a deep
long-slit spectrum obtained with the MSSSO 2.3-meter telescope confirm
this result and indicate that the emission in the filament is
LINER-like, based on the locations of the line ratios in diagnostic
diagrams (Fig. 19).  Dilute photoionization by a faint and distant AGN
with ionization parameter log~$U$ $\approx$ --3 (i.e.  the density of
ionizing photon is 10$^{-3}$ that of electron) could explain these
line ratios (e.g., Ferland \& Netzer 1983), although shock ionization
cannot be formally excluded (Dopita \& Sutherland 1995).

The kinematics of this filament can be used to distinguish between
these two scenarios. Multi-line imaging slightly shifted in velocity
space suggests that the line-emitting gas is blueshifted by $\sim$ 150
-- 200 km s$^{-1}$ with respect to systemic velocity. The MSSSO
spectrum of the filament confirms this small blueshift
(Fig. 20). These measurements indicate that the optical filament
coincides not only spatially with the H~I feature of Hameed et
al. (2001) but also kinematically.  Moreover, the narrow line widths
($\sim$ 80 -- 200 km s$^{-1}$) measured in the optical filament
indicate that the gas must not be affected significantly by shocks.
This strongly suggests that the line-emitting filament simply
represents tidal debris which are being ionized by the AGN in
NGC~7213. Only a small fraction of the H~I tidal complex is visible in
H$\alpha$ because the radiation field from NGC~7213 is not
isotropic. According to the unified model of Seyfert galaxies, our
line of sight to the Seyfert 1 nucleus has to lie within the opening
angle of the inner torus/accretion disk. Given the face-on orientation
of the host galaxy, this would imply that both our line of sight and
the line-emitting filament lie within the ionizing cone of the
accretion disk and above the galactic disk of NGC~7213.

\subsection{Circinus Galaxy}

The Circinus galaxy (Freeman et al. 1977) is host to a well-known
ionization cone and large-scale outflow visible in the radio (e.g.,
Elmouttie et al. 1998a), optical (e.g., Marconi et al. 1994; Lehnert
\& Heckman 1995; Veilleux \& Bland-Hawthorn 1997; Elmouttie et
al. 1998b; Wilson et al. 2000), near-infrared (Maiolino et al. 1998;
Ruiz et al. 2000), and X-rays (Sambruna et al. 2001a, 2001b; Smith \&
Wilson 2001).  The source of the ionization and outflow appears to be
the central Seyfert nucleus, although a circumnuclear starburst is
also present in this object (e.g., Veilleux \& Bland-Hawthorn 1997;
Elmouttie et al. 1998b)

A new TTF [O~III] $\lambda$5007 image of the Circinus galaxy is
presented in Figure 21. This image reaches fainter flux levels than
the Fabry-Perot data presented in Veilleux \& Bland-Hawthorn (1997).
The new data confirm the presence of the bright high-ionization
filaments within 0.5 kpc of the nucleus. But the deeper TTF image also
reveals unsuspected [O~III] emission which extends $\sim$ 1 kpc west
and north-west of the nucleus. Very faint, filamentary emission
appears to extend out to $\sim$ 1.2 kpc from the nucleus along
P.A. $\approx$ 315$^\circ$, but this needs to be confirmed with
deeper images. The TTF image also shows a few additional knots of
emission in the northern/northwestern quadrant and to the south. The
southernmost of these features appears to coincide with $Chandra$
source F in Smith \& Wilson (2001), located $\sim$ 500 pc due south from
the nucleus. This is shown in the lower left panel of Figure 21,
where the $Chandra$ data are superposed on the TTF [O~III] image.

\subsection{ESO484-G036}

ESO484-G036 is an edge-on Sb galaxy from the $IRAS$ Bright Galaxy
Sample (Soifer et al. 1989; F04335--2514).  The optical classification
of the nucleus of this galaxy is ambiguous due to weak H$\beta$ and
[O~III] emission (Kim et al. 1995; Veilleux et al. 1995). Long-slit
spectroscopy by Lehnert \& Heckman (1995) indicates the presence of
large positive gradients in [N~II]/H$\alpha$, [S~II]/H$\alpha$ and
[O~I]/H$\alpha$ along the minor axis of this galaxy. Conventional
H$\alpha$ + [N~II] narrowband imaging of this object by these same
authors reveals a cross-like structure aligned along the major and
minor axes of the disk.

The TTF data confirm the cross-like morphology of the line emission in
ESO484-G036 (especially visible in [N~II]; Fig. 22). The
[N~II]/H$\alpha$ ratio map also confirms the steep excitation gradient
along the minor axis of the galaxy. The disk emission presents a H~II
region-like [N~II]/H$\alpha$ ratios of $\sim$ 0.5 (but note that
[N~II]/H$\alpha$ reaches a minimum $\sim$ 1 kpc south-west of the
nucleus), while the emission 1 kpc above and below the disk have
[N~II]/H$\alpha$ that are in the range $\sim1-2.4$.  This line ratio
map is reminiscent of that of NGC~1482, but the larger distance of
ESO484-G036 ($\sim$ 60 Mpc instead of 20 Mpc for NGC~1482) makes the
extraplanar emission more difficult to resolve in this object. The new
TTF data add strong support to the idea that ESO484-G036 is indeed
host to a starburst-driven wind (Lehnert \& Heckman 1996), and also
show the promise of using excitation maps for systematic searches for
starburst-driven winds in the local universe (Veilleux \& Rupke 2002).

\subsection{\mr}

In the course of our TTF study, the radio-quiet quasar \mr\ was imaged
at H$\alpha$. This is the most distant ($z = 0.064$) object in the
sample. The results on this object have already been published by
Shopbell, Veilleux, \& Bland-Hawthorn (1999); the TTF images are
reproduced for the sake of completeness in Figure 23.  The TTF data
reveal a very extended nebula centered on and photoionized by this
quasar. The spiral-like complex is seen extending more or less
symmetrically over $\sim$ 200 kpc. Narrow-band images obtained at
slightly different wavelengths reveal a large-scale rotation pattern
which is in the opposite sense as that seen in the inner region of the
galaxy (see also Bergeron et al. 1983; N\"orgaard-Nielsen et
al. 1986). As discussed in detail in Shopbell et al., the large and
symmetric morphology of the gaseous envelope and its smooth
large-scale rotation suggest that the envelope did not originate with
a cooling flow, a past merger event, or an interaction with any of the
galaxies in the field. Shopbell et al. favor a model in which the
extended ionized nebula resides within a large complex of H~I gas
centered on the quasar. Slightly blueshifted ($\sim$ 300 km s$^{-1}$)
UV and warm X-ray absorbers are seen in HST and ASCA/ROSAT spectra of
\mr\ (Reynolds 1997; Komossa 2001; Monier et al. 2001; Ganguly,
Charlton, \& Eracleous 2001), but they are presumed to arise from
nuclear material near the AGN rather than from gas extending on
galactic scale. An error of a factor of 2.25 was recently found in the
flux calibration of the TTF data.  All fluxes listed in Shopbell et
al. (1999) should be scaled down by this factor. This correction does
not affect the conclusions of the paper.

\section{Discussion and Future Avenues of Research}

We have presented the results from a deep emission-line imaging survey
of 10 nearby starburst and active galaxies using the TTF on the AAT
and WHT. The topology and projected cross-section of the line-emitting
gas varies widely among the galaxies in the sample. Very large ($\ga$
80 kpc) ionized complexes are discovered around NGC~6240 and \mr.
Emission-line knots and wispy filaments are confirmed to be present in
the halo of NGC~4388 at distances of up to $\sim$ 30 kpc from the
active nucleus. Diffuse and filamentary gas associated with
large-scale (10 -- 20 kpc) ionization cones or outflows are detected
for the first time in NGC~1068, NGC~1365, and NGC~1482 and confirmed
in NGC~7213.  Emission-line structures on $\ga$ kpc scale are revealed
in the galactic winds of NGC~1705, Circinus galaxy, and ESO484-G036.

Multi-line imaging and long-slit spectroscopy of the gas found on
large scale reveal line ratios which are generally not H~II
region-like.  Shocks often contribute significantly to the ionization
of the outflowing gas on the outskirts of starburst galaxies. As
expected from shock models (e.g., Dopita \& Sutherland 1995), the
importance of shocks over photoionization by OB stars appears to scale
with the velocity of the outflowing gas (e.g., NGC~1482, NGC~6240, or
ESO484-G036 versus NGC~1705; NGC~3079 is an extreme example of a
shock-excited wind nebula; Veilleux et al. 1994), although other
factors like the starburst age, star formation rate, and the dynamical
state of the outflowing structure (e.g., pre- or post-blowout) must
also be important in determining the excitation properties of the gas
at these large radii (e.g., see discussion in Shopbell \&
Bland-Hawthorn 1998 and Veilleux \& Rupke 2002).  When an active
nucleus is present, the radiation field from the AGN appears to be the
primary source of ionization for the filaments within $\sim$ 10 -- 20
kpc of the nucleus, but the impact of the AGN tapers off at larger
distances (e.g., $\ga$ 30 kpc in NGC~4388), unless the source of
radiation is a powerful quasar (\mr).

Given the small sample size and methods of selection of the sample, no
statistical statement can be made on the frequency of occurrence of
large-scale nebulae around starburst and active galaxies in
general. The objects in our sample were specifically selected to host
ionization cones and/or galactic winds on $\sim$ kpc scale. The high
discovery rate of $\ga$ 10-kpc nebulae in our sample may not be
typical of the local population of starburst and active galaxies.  It
is also not clear at present whether the power and type of AGN (type 1
{\em versus} type 2) is important in determining the amount of ionized
gas detectable with the TTF. In the matter-bounded scenario (e.g.,
infinite reservoir of H~I gas surrounding the active nucleus in all
directions), one would expect a correlation between AGN ionizing power
and the detected mass of ionized material -- this may be why the
largest nebula is found around the most powerful AGN in our sample
(\mr). In this simple scenario, the orientation of the AGN accretion
disk to our line of sight (nearly face-on in type 1s {\em versus}
nearly edge-on in type 2s) should to first order determine the
geometry of the ionized material: biconical in Seyfert 2s (e.g.,
NGC~1068) and more isotropic in Seyfert 1s (e.g., \mr; see detailed
predictions in Mulchaey, Wilson, \& Tsvetanov 1996).  However, this
simple scenario clearly does not apply to all AGNs in our sample
(e.g., NGC~7213). The topology and projected cross-section of the warm
ionized gas in AGN necessarily also depend on the properties of the
host galaxy (e.g., morphological type, gas content) and immediate
environment (e.g, presence of a tidal complex as in the case of
NGC~7213).

Our pilot survey raises similar questions for starburst galaxies.
There is a clear need to expand the starburst sample to look for
possible dependence on the age and power of the starburst and the
dynamical state of the outflow.  These data will provide the material
to test the idea of using excitation maps (e.g., [N~II]/H$\alpha$) to
detect superwind galaxies (Veilleux \& Rupke 2002). If found to be
successful at low redshifts, this method could be used in the future
to more efficiently identify distant galaxies hosting powerful
starburst-driven winds.

The advent of tunable filters on 8-meter class telescopes [e.g.,
OSIRIS on the GranTeCan (Cepa et al. 2000) and the Maryland-Magellan
Tunable Filter on the Baade 6.5-m telescope] should improve the
sensitivity of emission-line galaxy surveys at least tenfold.
Measurements with this second generation of tunable filters will
provide direct quantitative constraints on the gaseous cross-section
of active and star-forming galaxies, and the importance of mass
exchange between galaxies and their environment. The detection of warm
ionized complexes that extend several tens of kpc over wide opening
angles would make them likely candidates for the higher column density
Ly$\alpha$ cloud population detected in quasar spectra (e.g., Bergeron
et al. 1994; Lanzetta et al. 1995; Norman et al. 1996; Steidel et
al. 2002). The kinematic information derived from these data will
constrain the origin of the gas: is it taking part in an outflow or is
it an extension of the HI disk, remnant accreting gas from galaxy
formation, or debris from a recent galaxy interaction or ram-pressure
stripping episode?

%\clearpage

\acknowledgements

We are grateful to E. Brinks for providing the H~I map of NGC~1068
prior to publication and to K. Iwasawa and A. Wilson for the $Chandra$
X-ray data of NGC~4388. We thank the referee, Matthias Ehle, for
constructive comments. This article was written while S.V.  was on
sabbatical at the California Institute of Technology and the
Observatories of the Carnegie Institution of Washington; S.V. thanks
both of these institutions for their hospitality.  S.V. acknowledges
partial support of this research by a Cottrell Scholarship awarded by
the Research Corporation, NASA/LTSA grant NAG 56547, and NSF/CAREER
grant AST-9874973.  D.S.R. was also supported in part by NSF/CAREER
grant AST-9874973.  This research has made use of the NASA/IPAC
Extragalactic Database (NED)), which is operated by the Jet Propulsion
Laboratory, California Institute of Technology, under contract with
the National Aeronautics and Space Administration.

%\clearpage

\clearpage

\begin{figure}
%\centerline{\epsfig{figure=veilleux.fig1.eps,width=1.2\textwidth,angle=270}}
\caption{ NGC~1068 in ($a$) R0 continuum, ($b$) H$\alpha$, ($c$)
[O~III] $\lambda$5007, and ($d$) [O~III] $\lambda$5007/H$\beta$ ratio
($\equiv$ 2.85 x [O~III] $\lambda$5007/H$\alpha$). North is up and
east to the left. The cross in each panel marks the location of
the nucleus from NED. The flux scale of the emission-line maps is
logarithmic, while the ratio map is on a linear scale.  The flux scale
for the [O~III] data ranges from --17.5 to --15.0, while the scale for
the H$\alpha$ map ranges from --18.5 to --15.0. Emission from
both [O~III] and H$\alpha$ is detected for the first time out to
$\sim$ 11 kpc from the nucleus in the north-east (upper left) and
south-west (lower right) quadrants, roughly aligned with the
ionization cone on smaller scale. Note the high [O~III]/H$\beta$
ratios in the large-scale filaments.  The diffuse emission features at
the eastern edge of panel ($b$) and southern edge of panel ($c$) are
artifacts of reflective ghosts associated with the TTF. }
\end{figure}
 
\begin{figure}
%\centerline{\epsfig{figure=veilleux.fig2.eps,width=0.5\textwidth,angle=270}}
\caption{ Multiwavelength comparison in NGC~1068. ($a$) H$\alpha$ and
H~I 21-cm contour map (from Brinks 2003; private communication) and
($b$) [O~III] $\lambda$5007 and X-ray contour map from Young, Wilson,
\& Shopbell (2001). North is up and east to the left. The
emission-line maps are on the same flux scale as in Figure 1. The
contours in the HI map are at 25, 100, 150, 200, and 250 Jansky
s$^{-1}$ beam$^{-1}$. Refer to Young et al. (2001) for the values of
the X-ray contour levels. The cross in each panel marks the location
of the nucleus from NED. Note the slight misalignment between the
large-scale filaments and the inner ionization cone defined by the
X-ray emission.  The diffuse emission feature at the southern edge of
panel ($b$) is a reflective ghost associated with the TTF.}
\end{figure}

\begin{figure}
\centerline{\epsfig{figure=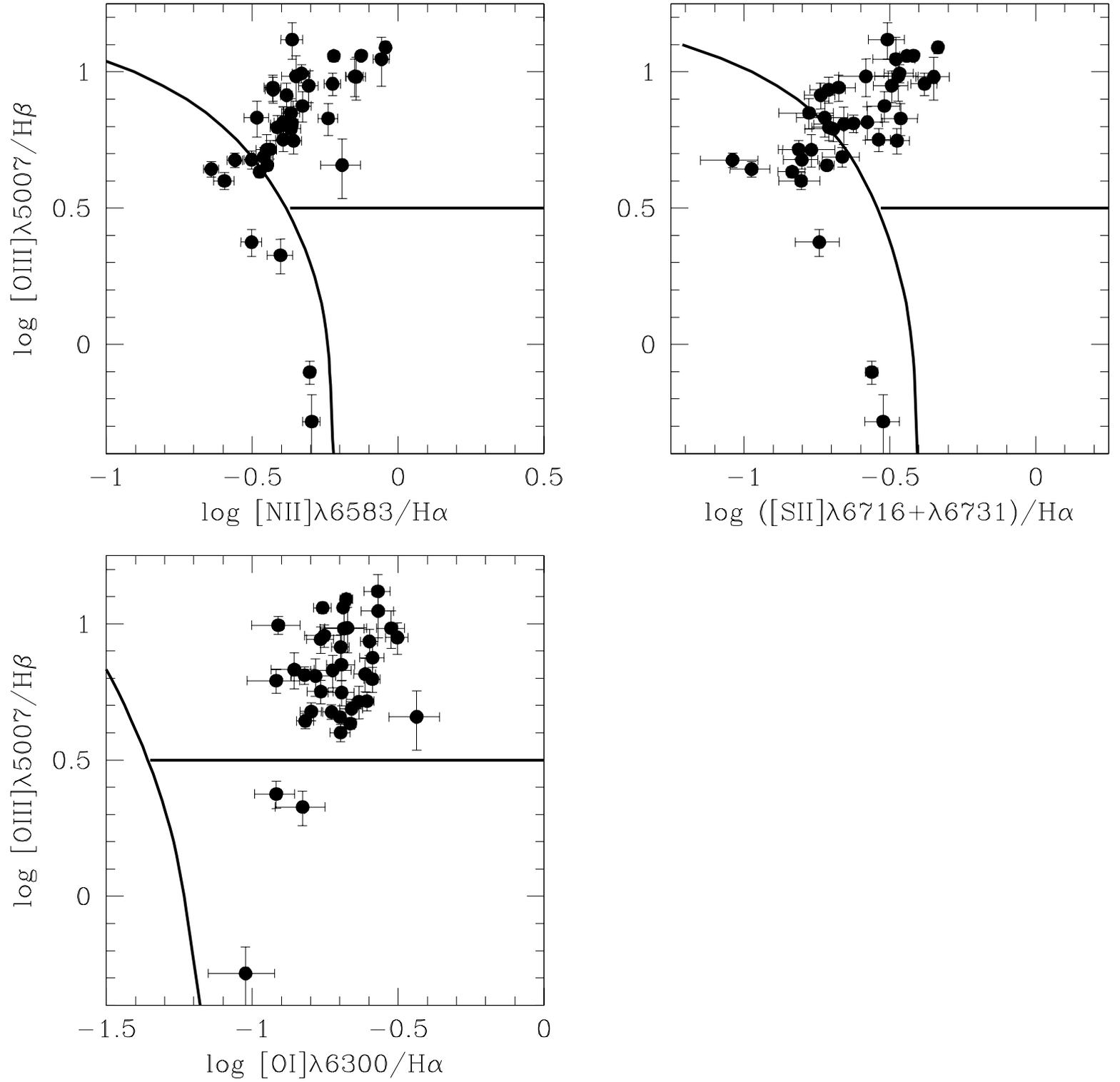,width=1.2\textwidth,angle=0}}
\caption{ Emission-line ratios measured along the north-east filament
(P.A. = 3$^\circ$) located at $\sim$ 6 -- 12 kpc from the nucleus of
NGC~1068. Most of the line ratios are Seyfert-like, therefore
indicating that the filament is ionized by the central AGN.  }
\end{figure}

\begin{figure}
%\centerline{\epsfig{figure=veilleux.fig4.eps,width=0.8\textwidth,angle=270}}
\caption{ Continuum-subtracted long-slit spectrum obtained along the
north-east filament located at $\sim6-12$ kpc from the nucleus of
NGC~1068. The slit is oriented along P.A. = 3$^\circ$ with North at the
top. The white horizontal band is hiding the spectrum of a bright
star. The flux scale is logarithmic and in units of erg s$^{-1}$
cm$^{-2}$.  The velocities are relative to systemic (= 1792 km
s$^{-1}$). The blueshift, velocity gradient, and narrow line widths in
the brighter portion of the filament are consistent with quiescent gas
in galactic rotation.  Note, however, that the emission-line profiles
become broader in the fainter (southern) portion of the filament,
perhaps a sign that shocks become important here or the gas ionized by
the AGN spans a broader range of line-of-sight velocities.}
\end{figure}

\begin{figure}
%\centerline{\epsfig{figure=veilleux.fig5.eps,width=1.2\textwidth,angle=270}}
\caption{ Outer regions of NGC~1365 in ($a$) B4 continuum, ($b$)
H$\alpha$, ($c$) [O~III] $\lambda$5007, and ($d$) [O~III]
$\lambda$5007/H$\beta$ ratio ($\equiv$ 2.85 $\times$ [O~III]
$\lambda$5007/H$\alpha$). North is up and east to the left. The cross
in each panel marks the location of the red continuum peak. The
flux scale of the emission-line maps is logarithmic, while the ratio
map is on a linear scale. A $\sim$ 7-kpc ionization cone is visible in
the ratio maps.  }
\end{figure}
 
\begin{figure}
%\centerline{\epsfig{figure=veilleux.fig6.eps,width=1.2\textwidth,angle=270}}
\caption{ Outer regions of NGC~1365 in ($a$) R0 continuum, ($b$)
H$\alpha$, ($c$) [N~II] $\lambda$6583, and ($d$) [N~II]
$\lambda$6583/H$\alpha$ ratio.  North is up and east to the left. The
cross in each panel marks the location of the red continuum
peak. The flux scale of the emission-line maps is logarithmic,
while the ratio map is on a linear scale. A $\sim$ 12-kpc boxy region
with [N~II]/H$\alpha$ $>$ 1 is detected around the nucleus.}
\end{figure}

\begin{figure}
%\centerline{\epsfig{figure=veilleux.fig7.eps,width=1.2\textwidth,angle=270}}
\caption{ Central regions of NGC~1365 in ($a$) B4 continuum, ($b$)
H$\alpha$, ($c$) [O~III] $\lambda$5007, and ($d$) [O~III]
$\lambda$5007/H$\beta$ ratio ($\equiv$ 2.85 x [O~III]
$\lambda$5007/H$\alpha$). North is up and east to the left. The cross
in each panel marks the location of the red continuum peak. The
flux scale of the emission-line maps is logarithmic, while the ratio
map is on a linear scale. The ratio map clearly shows a ionization
cone bisected by a dust lane.  }
\end{figure}
 
\begin{figure}
%\centerline{\epsfig{figure=veilleux.fig8.eps,width=1.2\textwidth,angle=270}}
\caption{ Central regions of NGC~1365 in ($a$) R0 continuum, ($b$)
H$\alpha$, ($c$) [N~II] $\lambda$6583, and ($d$) [N~II]
$\lambda$6583/H$\alpha$ ratio.  The contours in ($d$) show the [O~III]
$\lambda$5007/H$\beta$ ratios from Fig. 7. North is up and east to the
left. The cross in each panel marks the location of the red continuum
peak. The flux scale of the emission-line maps is logarithmic, while
the ratio map is on a linear scale. The contour levels of the
[O~III]/H$\beta$ ratio map are at 0.58, 2.9, and 9.2 $\times$
10$^{-15}$ erg s$^{-1}$ cm$^{-2}$ arcsecond$^{-2}$. A large boxy
region with [N~II]/H$\alpha$ $>$ 1 is detected around the nucleus. }
\end{figure}

\begin{figure}
%\centerline{\epsfig{figure=veilleux.fig9.eps,width=1.2\textwidth,angle=270}}
\caption{ NGC~1482 in ($a$) R0 continuum, ($b$) H$\alpha$, ($c$)
[N~II] $\lambda$6583 and $Chandra$ X-ray contour map, and ($d$) [N~II]
$\lambda$6583/H$\alpha$ ratio.  North is up and east to the left. The
crosses in each panel indicate the locations of the two peaks in the
red continuum. The flux scale of the emission-line maps is
logarithmic, while the ratio map is on a linear scale. The contour
levels of the X-ray map are at 3.0, 5.4, 17, and 169 $\times$
10$^{-4}$ counts s$^{-1}$ arcsecond$^{-2}$ (based on an effective
exposure time of 24.40 ksec). Note the tight match between the X-ray
emission and some of the optical filaments.  }
\end{figure}

\begin{figure}
%\centerline{\epsfig{figure=veilleux.fig10.eps,width=1.2\textwidth,angle=270}}
\caption{ $Chandra$ X-ray contour map superposed on a very deep
H$\alpha$ + [N~II] $\lambda$6583 image of NGC~1482. North is up and
east to the left. The crosses indicate the locations of the two peaks
in the red continuum. The flux scale is logarithmic.  The contour levels
of the X-ray map are the same as in Figure 9. Filamentary line
emission is detected out to $\sim$ 7 kpc north-west of the nucleus,
and perhaps out to $\sim$ 12 kpc to the north-east.  Diffuse emission
may also be present $\sim$ 10 kpc south of the nucleus. A bright star
on the eastern edge of the image was masked for display purposes; it may 
be responsible for reflective ghosts that can be confused with 
faint emission-line features. }
\end{figure}

\begin{figure}
%\centerline{\epsfig{figure=veilleux.fig11.eps,width=0.8\textwidth,angle=270}}
\caption{ Long-slit spectra of NGC~1482 centered on H$\alpha$ and
[N~II] $\lambda$6583. The position angle of the slit in both spectra
is perpendicular to the galaxy disk (North is up). In the left two
panels, the slit is centered on the nucleus while it is offset
5$\arcsec$ west of the nucleus in the two panels on the right.  The
velocities are relative to systemic (= 1915 km s$^{-1}$). The flux
scale is logarithmic to emphasize the presence of the faint
emission-line knot (indicated by the black circle) at $\sim$ 3 kpc
North and +400 km s$^{-1}$, more easily visible in [N~II] than at
H$\alpha$. }
\end{figure}

\begin{figure}
%\centerline{\epsfig{figure=veilleux.fig12.eps,width=1.2\textwidth,angle=270}}
\caption{ NGC~1705 in ($a$) R0 continuum, ($b$) H$\alpha$, ($c$)
[N~II] $\lambda$6583, and ($d$) [N~II] $\lambda$6583/H$\alpha$ ratio.
North is up and east to the left. The cross in each panel marks
the location of the red continuum peak.  The flux scale of the
emission-line maps is logarithmic, while the ratio map is on a linear
scale. Note the well-known arcs and bubble-like structures in this
galaxy, and the surprising lack of structure in the excitation map
across these features. }
\end{figure}

\begin{figure}
%\centerline{\epsfig{figure=veilleux.fig13.eps,width=1.2\textwidth,angle=270}}
\caption{ Inner regions of NGC~4388 in ($a$) R0 continuum, ($b$)
H$\alpha$, ($c$) [N~II] $\lambda$6583 and $Chandra$ X-ray contour map
from Iwasawa et al. (2003), and ($d$) [N~II] $\lambda$6583/H$\alpha$
ratio.  North is up and east to the left. The cross in each panel
marks the location of the radio continuum peak from Falcke, Wilson, \&
Simpson (1998). The flux scale of the emission-line maps is
logarithmic, while the ratio map is on a linear scale. Refer to
Iwasawa et al. (2003) for the values of the X-ray contour levels and
for more detail on the X-ray data. Note the good match between the
X-ray emission and the emission-line filaments above the disk of the
galaxy, and the modest [N~II]/H$\alpha$ ratios ($\la$ 0.7) in the
extraplanar material.}
\end{figure}

\begin{figure}
%\centerline{\epsfig{figure=veilleux.fig14.eps,width=1\textwidth,angle=270}}
\caption{ Outer regions of NGC~4388 in ($a$) R0 continuum, ($b$)
H$\alpha$, ($c$) [N~II] $\lambda$6583, and ($d$) [N~II]
$\lambda$6583/H$\alpha$ ratio.  North is up and east to the left. The
cross in each panel marks the location of the radio continuum
peak from Falcke, Wilson, \& Simpson (1998). The flux scale of the
emission-line maps is logarithmic, while the ratio map is on a linear
scale. Note the modest [N~II]/H$\alpha$ ratios ($\la$ 0.7) in the
outer filaments.  }
\end{figure}

\begin{figure}
\centerline{\epsfig{figure=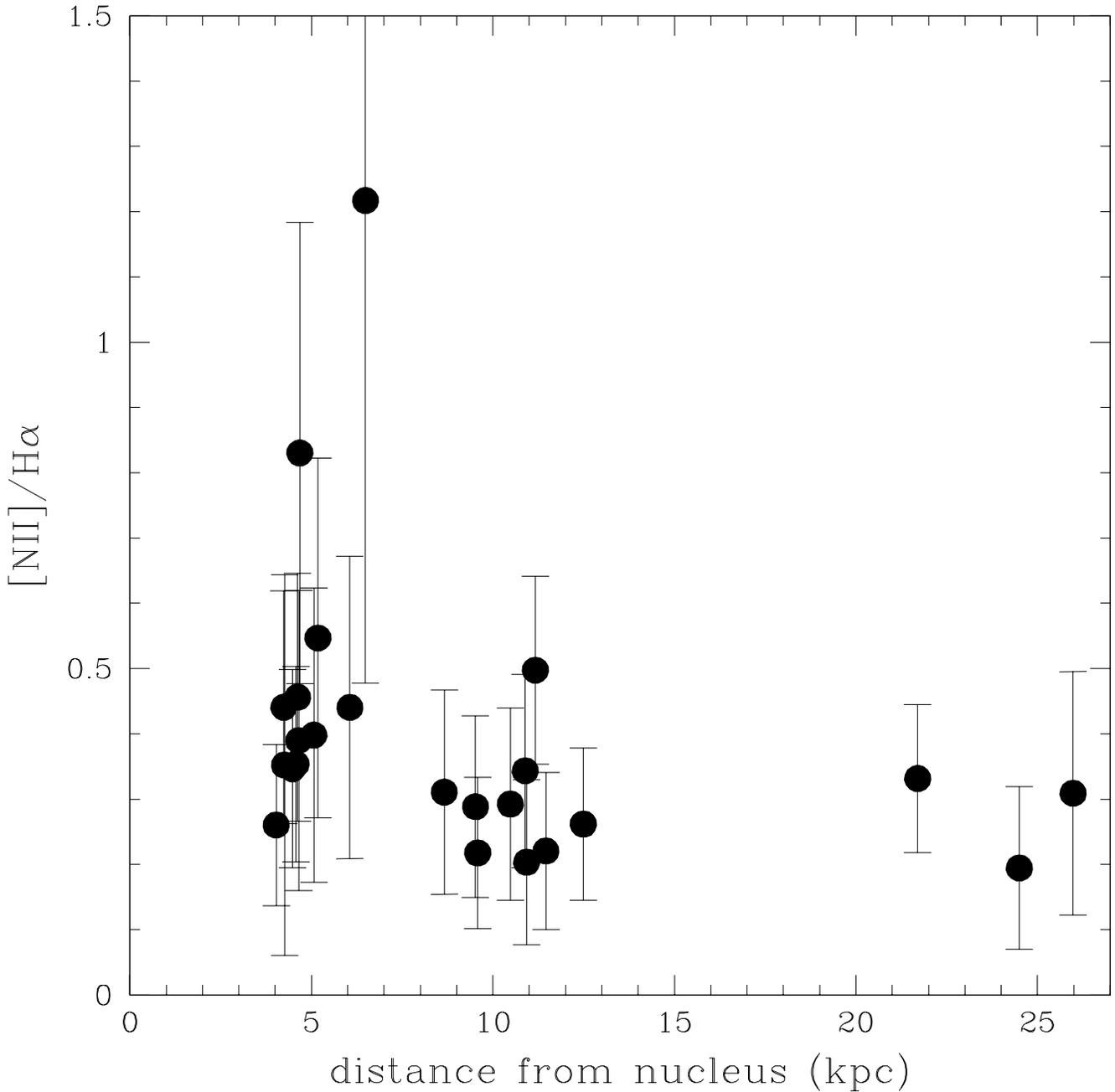,width=1\textwidth,angle=0}}
\caption{ [N~II] $\lambda$6583/H$\alpha$ ratios as a function of
galactocentric distance in NGC~4388. A 5-pixel boxcar kernel was used
to smooth the [N~II]/H$\alpha$ ratio map. The values shown in the
figure represent the mean and $\pm$ 1 $\sigma$ around the mean
calculated over regions containing from 50 to 350 pixels. The
[N~II]/H$\alpha$ ratios beyond $\sim$ 7 kpc generally stay below
$\sim$ 0.7 and show no obvious gradient with distance from the nucleus. }
\end{figure}

\begin{figure} 
%\centerline{\epsfig{figure=veilleux.fig16.eps,width=1.2\textwidth,angle=270}}
\caption{ Outer regions of NGC~6240 in ($a$) R0 continuum, ($b$)
H$\alpha$, ($c$) [N~II] $\lambda$6583 and $Chandra$ X-ray contour map
from Komossa et al. (2003), and ($d$) [N~II] $\lambda$6583/H$\alpha$
ratio.  North is up and east to the left. The crosses indicate the
positions of the binary AGN in this object. The flux scale of the
emission-line maps is logarithmic, while the ratio map is on a linear
scale.  The contour levels of the X-ray map are at 7.0, 14, 28, and
111 $\times$ 10$^{-4}$ counts s$^{-1}$ arcsec$^{-2}$ (based on an
effective exposure time of 37 ksec). Complex filamentary emission
extends over 70 $\times$ 80 kpc. Note the good match between the X-ray
emission and the high-[N~II]/H$\alpha$ loop and filament at $R \approx
15 - 25$ kpc and P.A. $\approx$ 270$^\circ$.  These features are
probably associated with a galactic outflow.  }
\end{figure}

\begin{figure} 
%\centerline{\epsfig{figure=veilleux.fig17.eps,width=1.2\textwidth,angle=270}}
\caption{ Inner regions of NGC~6240 in ($a$) R0 continuum, ($b$)
H$\alpha$, ($c$) [N~II] $\lambda$6583 and $Chandra$ X-ray contour map
from Komossa et al. (2003), and ($d$) [N~II] $\lambda$6583/H$\alpha$
ratio.  North is up and east to the left. The crosses indicate the
positions of the binary AGN in this object. The flux scale of the
emission-line maps is logarithmic, while the ratio map is on a linear
scale.  The contour levels of the X-ray map are at 3.3, 10, 30, and 60
$\times$ 10$^{-4}$ counts s$^{-1}$ arcsec$^{-2}$ (based on an
effective exposure time of 37 ksec). Note the good match between the
brighter portions of the 5-kpc X-ray loop and a high-[N~II]/H$\alpha$
feature at $R \approx 5$ kpc and P.A. $\approx$ 290$^\circ$.  Both
features are probably associated with a galactic outflow.}
\end{figure}

\begin{figure}
%\centerline{\epsfig{figure=veilleux.fig18.eps,width=1.2\textwidth,angle=270}}
\caption{ NGC~7213 in ($a$) the R0 continuum, ($b$) H$\alpha$, ($c$)
[N~II] $\lambda$6583, and ($d$) [N~II] $\lambda$6583/H$\alpha$ ratio.
North is up and east to the left. The cross in each panel marks
the location of the red continuum peak. The flux scale of the
emission-line maps is logarithmic, while the ratio map is on a linear
scale. A line-emitting filament is detected 19 kpc south-west of the
nucleus, well beyond the optical extent of NGC~7213. The existence of
this filament has been independently confirmed by Hameed et
al. (2001). The broad diffuse emission off-centered from the nucleus
in the TTF data is a reflective ghost. }
\end{figure}

\clearpage

\begin{figure}
\centerline{\epsfig{figure=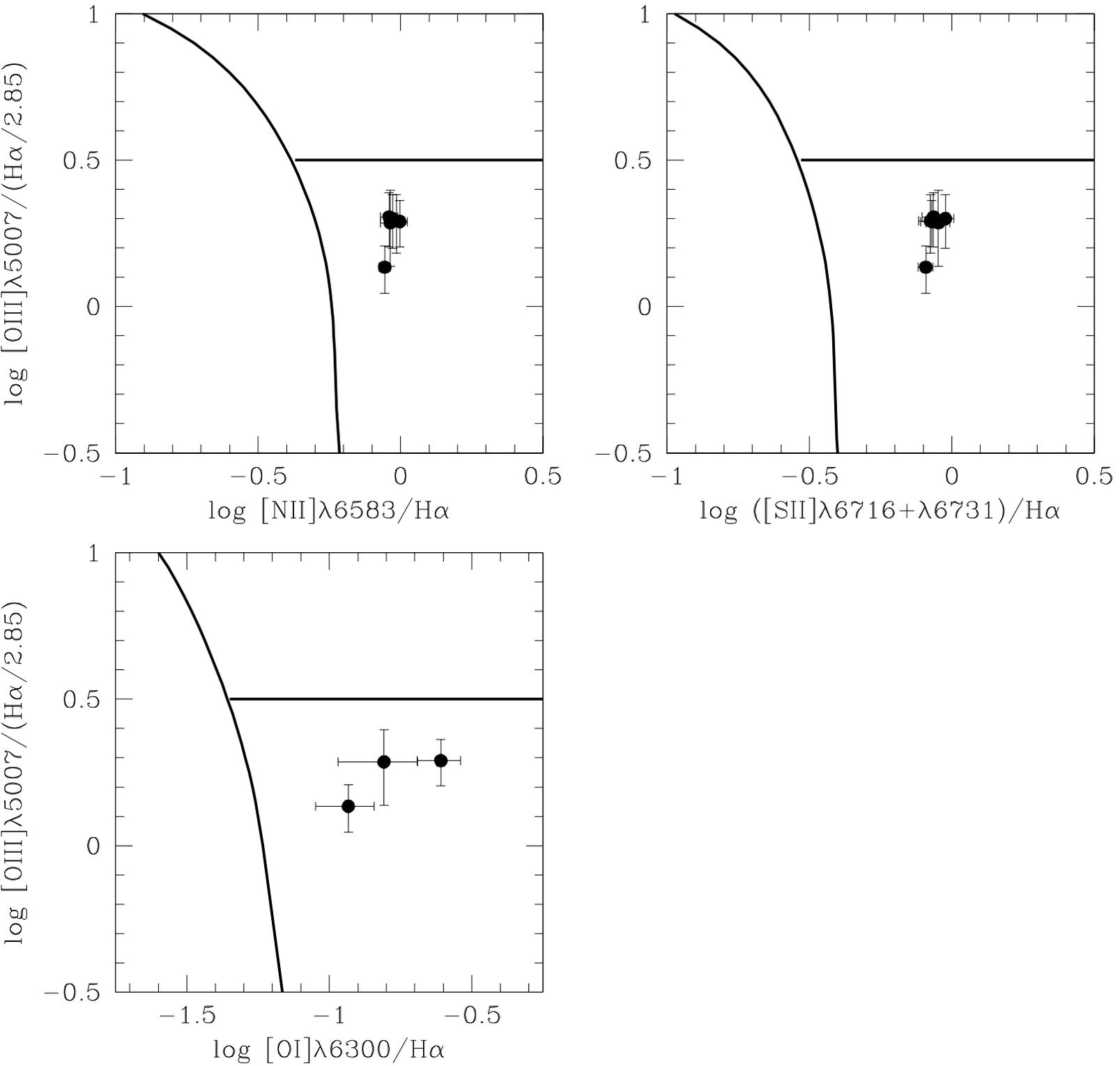,width=1.2\textwidth,angle=0}}
\caption{ Emission-line ratios measured in the southern filament of
NGC~7213. All of the line ratios are LINER-like. Given the narrow
widths of the emission lines in the filament (see Fig. 20), this
suggests that the filament is ionized by the diluted radiation from
the central AGN rather than by shocks.  }
\end{figure}

\begin{figure}
%\centerline{\epsfig{figure=veilleux.fig20.eps,width=0.8\textwidth,angle=270}}
\caption{ Continuum-subtracted long-slit spectrum obtained along the
southern filament at $R \approx$ 19 kpc and P.A. $\approx$
110$^\circ$. The flux scale for [O~III] is linear and in units of erg
s$^{-1}$ cm$^{-2}$, while it is logarithmic for the other lines.
The gas is blueshifted by $\sim$ 200 km s$^{-1}$ with respect to the
systemic velocity, 1148 km s$^{-1}$. This velocity coincides with the
neutral tidal gas detected by Hameed et al. (2001) at that same
position.  }
\end{figure}

\begin{figure}
%\centerline{\epsfig{figure=veilleux.fig21.eps,width=1.2\textwidth,angle=270}}
\caption{ Circinus galaxy in ($a$) the 2MASS K-band continuum, ($b$)
high-resolution H$\alpha$ map obtained with $HST$ (Wilson et
al. 2000), ($c$) [O~III] $\lambda$5007, and ($d$) $Chandra$ X-ray
contour map from Smith \& Wilson (2001) superposed on the [O~III] map.
North is up and east to the left. The cross in each panel marks the
location of the radio/X-ray nucleus of Smith \& Wilson (2001). The
flux scale of the emission-line maps is logarithmic, while the ratio
map is on a linear scale. Refer to Smith \& Wilson (2001) for the
values of the X-ray contour levels. The TTF reach a slightly fainter
flux limit than the data of Veilleux \& Bland-Hawthorn (1997) and
reveal new emission-line features $\ga$ 1 kpc west and north-west
from the nucleus.}
\end{figure}

\begin{figure}
%\centerline{\epsfig{figure=veilleux.fig22.eps,width=1.2\textwidth,angle=270}}
\caption{ ESO484-G036 in ($a$) the R0 continuum, ($b$) H$\alpha$,
($c$) [N~II] $\lambda$6583, and ($d$) [N~II] $\lambda$6583/H$\alpha$
ratio.  North is up and east to the left. The cross in each panel
marks the location of the red continuum peak. The flux scale of
the emission-line map is logarithmic, while the ratio map is on a
linear scale. Note the cross-like structure in ($c$) and ($d$), the
high-[N~II]/H$\alpha$ extraplanar material, and the [N~II]/H$\alpha$
minimum $\sim$ 1 kpc south-west from the nucleus. }
\end{figure}

\begin{figure}
%\centerline{\epsfig{figure=veilleux.fig23.eps,width=1.0\textwidth,angle=0}}
\caption{Deep H$\alpha$ and I-band images of the field surrounding the
quasar \mr, reproduced from Shopbell, Veilleux, \& Bland-Hawthorn
(1999).  Panels ($a$) and ($b$) are 1200-second exposures at redshifts of
0.0640 and 0.0645, respectively, panel ($c$) is an I-band continuum
image of the same field, and panel ($d$) is a summed H$\alpha$
image. North is up and east to the left. The flux scale is
logarithmic.  A bright star (S), a nearby cluster galaxy (G1), and a
number of emission-line knots from Macchetto et al. (1990) have been
labeled. The emission is detected on a scale of $\sim$ 200 kpc.}
\end{figure}
 
\clearpage

\begin{deluxetable}{llll}
\tablecaption{Sample}
\tablewidth{0pt}
\tablehead{
\colhead{Name} & \colhead{Redshift\tablenotemark{a}} & \colhead{Nuclear Type\tablenotemark{b}} & \colhead{Main Findings}}
\startdata 
NGC 1068	& 0.0038 & AGN           & Ionization cone detected over $\sim$ 20-kpc scale \\
NGC 1365	& 0.0055 & AGN           & Ionization cone detected over $\sim$ 12-kpc scale \\
NGC 1482	& 0.0064 & Starburst     & Galactic wind detected over $\sim$ 20-kpc scale  \\
NGC 1705	& 0.0021 & Starburst     & Featureless [N~II]/H$\alpha$ map of $\sim$ 3-kpc wind region\\
NGC 4388	& 0.0084 & AGN           & Complex of filaments detected over $\sim$ 35-kpc scale \\
NGC 6240	& 0.0245 & AGN/Starburst & Complex of loops and filaments extending $\sim$ 70 $\times$ 80 kpc\\
NGC 7213	& 0.0060 & AGN           & Ionized tidal filament at $\sim$ 19 kpc from nucleus \\
Circinus Galaxy	& 0.0015 & AGN/Starburst & Highly ionized filaments out to $\sim$ 1.2 kpc from nucleus\\ 
ESO~484-G036	& 0.0162 & Starburst     & Cross-like [N~II]/H$\alpha$ map indicative of wind\\
\mr	        & 0.0640 & QSO           & Spiral-like nebula detected over $\sim$ 200-kpc scale\\
\enddata
\tablenotetext{a}{The redshifts are from the NASA Extragalactic Database (NED).}
\tablenotetext{b}{See text for references to the nuclear spectral type.}
\end{deluxetable}

\begin{deluxetable}{clccccc}
%\rotate
%\tabletypesize{\scriptsize}
\tablecaption{Journal of Observations \label{obs_tab}}
\tablewidth{0pt}
\tablehead{
\colhead{Run} & \colhead{UT dates} & \colhead{Telescope} & \colhead{Instrument} & \colhead{CCD} &\colhead{$\%$ Dark} &\colhead{Seeing}\\
\colhead{(1)} & \colhead{(2)} & \colhead{(3)} & \colhead{(4)} & \colhead{(5)} & 
\colhead{(6)} & \colhead{(7)}
}
\startdata
1i&	1998 Aug 30, Sep 03&	AAT	&	RTTF&	MITLL2	&14, 47	&1\farcs3\\
2i&	1999 Mar 04-06	&	WHT	&	RTTF&	TEK2	&14	&0\farcs6-1\farcs8\\
3i&     2000 Feb 05-07, 14-16&  AAT     &       BTTF&   MITLL3  &22, 99 &1\farcs5\\
4i&	1999 Nov 28	&	AAT	&	RTTF&	MITLL2	&43	&1\farcs0-1\farcs5\\
5i&	2000 Feb 17-20	&	AAT	&	BTTF&	MITLL3	&0-6	&1\farcs3-2\farcs8\\
6i&	2000 Jul 28	&	AAT	&	RTTF&	MITLL2A	&91	&1\farcs3-1\farcs6\\
7i&	2000 Dec 15-17	&	AAT	&	BTTF&	MITLL2A	&24-45	&1\farcs5-2\farcs5\\
8i&	2002 May 16-20	&	AAT	&	BTTF&	EEV2	&41-83	&1\farcs5-4\farcs0\\
  &  \\
1s&	2001 Sep 18-20	&	MSSSO 2.3m&	DBS&	SITe	&86-98	&1\farcs4-1\farcs5\\
2s&	2001 Dec 19	&	AAT	&	RGO 25cm&EEV2	&79	&1\farcs7	\\
\enddata

\tablecomments{ Col. (1): Observing run number. Suffix ``i'' refers to
imaging data, while suffix ``s'' refers to long-slit spectroscopy.
(2): Dates of observations.  Col. (3): Telescope used for the
observations. AAT = Anglo-Australian 3.9-meter Telescope, WHT =
William Herschel 4.2-meter Telescope, and MSSSO 2.3m = Mount Stromlo
-- Siding Springs 2.3-meter Telescope.  Col.(4): Instrument used for
the observations. RTTF = Red Taurus Tunable Filter, BTTF = Blue Taurus
Tunable Filter, DBS = Double-Beam Spectrograph, and RGO 25cm = Royal
Greenwich Observatory spectrograph equipped with 25-cm
camera. Col.(5): CCD used for the observations. MITLL2/MITLL2A = 2048
x 4096 x 15 $\mu$m, MITLL3 = 2048 x 4096 x 15 $\mu$m (deep depletion),
TEK2 = 1024 x 1024 x 24 $\mu$m, EEV2 = 2048 x 4096 x 13 $\mu$m (blue
sensitive), and SITe = 1752 x 532 x 15 $\mu$m. Col.(6): Phase of the
Moon during the observations. 100\% = New Moon. Col.(7): Seeing 
in arcseconds during the night.}
\end{deluxetable}

\begin{deluxetable}{lccccccclccl}
%\rotate
%\tabletypesize{\scriptsize}
\tiny
\tablecaption{Imaging Observations\label{im_tab}}
\tablewidth{0pt}
\tablehead{
\colhead{Object} & \colhead{Run} & \colhead{Line} & \colhead{Filter/Tilt} & \colhead{$t_{exp}$} &\colhead{$\lambda_{obs}$}   &\colhead{$\lambda_{exp}$} &\colhead{$\lambda_{cont}$} &\colhead{$\Delta\lambda_{eff}$} &\colhead{CF/FS?} &\colhead{Straddle?} &\colhead{Sens.} \\
\colhead{(1)} & \colhead{(2)} & \colhead{(3)} & \colhead{(4)} & \colhead{(5)} & 
\colhead{(6)} & \colhead{(7)} & \colhead{(8)} & \colhead{(9)} & \colhead{(10)} & \colhead{(11)} & \colhead{(12)}
}
\startdata
NGC 1068	&2i	&H$\alpha$&I0/10	&3600		&6588	&6588	&(I0/I1)	&11.0		&N &N &\phn3.0\\
                &3i	&[OIII]	&B4/\phn0	&3663		&5028	&5026	&4908		&17.1		&Y &N &15	\\
NGC 1365	&7i	&[OIII]	&B4/\phn0	&2$\times$976	&5034	&5034	&4994,5075	&20.5		&Y &Y &69	\\
		&	&H$\alpha$&R0/15	&1$\times$976	&6600	&6599	&6554,6647	&13.9		&Y &Y &\phn8.5\\
		&	&[NII]	&R0/15		&1$\times$976	&6620	&6619	&6555,6647	&15.7		&Y &Y &16  	\\
NGC 1482	&7i	&H$\alpha$&R0/15	&2$\times$976	&6606	&6605	&6561,6653	&14.0		&Y &Y &\phn5.3\\
		&	&[NII]	&R0/15		&2$\times$976	&6627	&6626	&6562,6664	&13.9		&Y &Y &\phn4.5\\
NGC 1705	&7i	&H$\alpha$&R0/15	&3$\times$976	&6578	&6577	&6533,6625	&16.8		&Y &Y &\phn8.3\\
		&	&[NII]	&R0/15		&2$\times$976	&6597	&6598	&6533,6625	&16.7		&Y &Y &\phn8.4\\
NGC 4388	&8i	&H$\alpha$&R0/15	&4$\times$976	&6605	&6618	&6551,6652	&15.5		&Y &Y &\phn5.2\\
		&	&[NII]	&R0/15		&6$\times$976	&6628	&6639	&6554,6655	&17.1		&Y &Y &\phn4.6\\
NGC 6240	&8i	&H$\alpha$&R0/\phn0	&6$\times$976	&6724	&6723	&6670,6772	&15.5		&Y &Y &\phn4.0\\
		&	&[NII]	&R0/\phn0	&4$\times$976	&6743	&6745	&6669,6770	&16.6		&Y &Y &\phn5.7\\
NGC 7213	&4i	&H$\alpha$/[NII]&R0/\phn0	&6$\times$611	&\tablenotemark{a}&6602,6623&\tablenotemark{a}&27.1	&Y &N &12\\
		&6i	&H$\alpha$/[NII]&R0/\phn0	&6$\times$300	&\tablenotemark{b}&6602,6623&\tablenotemark{b}&\phn8.3	&N &N &10\\
Circinus Galaxy	&5i	&[OIII]	&B4/\phn0	&6$\times$976	&5017	&5014	&4987,5048	&19.5		&Y &Y &39 	\\ 
ESO 484-G036	&7i	&H$\alpha$&R0/\phn0	&1$\times$976	&6670	&6669	&6624,6718	&15.9		&Y &Y &12  	\\
		&	&[NII]	&R0/\phn0	&1$\times$976	&6692	&6690	&6624,6719	&15.9		&Y &Y &11  	\\
\mr	        &1i	&H$\alpha$&R1/16	&2$\times$600	&6983	&6983	&(LDSS I)	&12.4		&N &N &\phn2.5	\\
	        &	&H$\alpha$&R1/16	&2$\times$600	&6986	&6983	&(LDSS I)	&12.4		&N &N &\phn2.5	\\
\enddata

\tablecomments{ Col. (1): Name of target. Col. (2): Run number (see
Table 2). Col. (3): Emission line. [NII] = [N~II] $\lambda$6583,
[OIII] = [O~III] $\lambda$5007. Col.(4): Filter used and tilt in
degrees.  Col.(5): Total exposure time, in seconds.  Col.(6-8):
Observed, expected (based on the galaxy's systemic velocity), and
continuum wavelengths at the center of each object.  Col.(9):
Effective bandpass in \AA.  Col. (10): Use of charge shuffling and
frequency switching technique (see \S 2). Col. (11): Use of the
wavelength straddling mode (see \S 2). Col.(10): Sensitivity ($\sim$
1-$\sigma$) of observations, in units of $10^{-18}$ erg s$^{-1}$
cm$^{-2}$ arcsecond$^{-2}$.  } 

\tablenotetext{a}{For run 4i, NGC 7213 was observed at 6 different
wavelengths, 5 of which contain some line emission: 6585, 6594, 6603,
6625, 6634, and 6643 (all in \AA, at the location of the line-emitting
filament).}

\tablenotetext{b}{For run 6i, NGC 7213 was observed at 6 different
wavelengths, 5 of which contain some line emission: 6594, 6615, 6617,
6620, 6626, and 6634 (all in \AA, at the location of the line-emitting
filament).}
\end{deluxetable}

\begin{deluxetable}{lcrlrl}
%\rotate
%\tabletypesize{\scriptsize}
\tablecaption{Spectroscopic Observations \label{spec_tab}}
\tablewidth{0pt}
\tablehead{
\colhead{Object} & \colhead{Run} & \colhead{$t_{exp}$} & \colhead{Slit} &\colhead{PA} &\colhead{description}\\
\colhead{(1)} & \colhead{(2)} & \colhead{(3)} & \colhead{(4)} & \colhead{(5)} & 
\colhead{(6)}
}
\startdata
NGC 1068	&1s	 &8$\times$1800	&2\arcsec$\times$6\farcm7	  &6	&along filament in NE ionization cone	\\
		&2s	 &3$\times$1800	&2\arcsec$\times$100\arcsec	  &3	&along filament in NE ionization cone	\\
NGC 1482	&1s	 &15$\times$1200&2\arcsec$\times$6\farcm7	&103	&15\arcsec\ North - 16\arcsec\ South of disk	\\
		&	 &1$\times$ 600	&2\arcsec$\times$6\farcm7	&103	&5\arcsec\ South of disk		\\
		&	 &1$\times$ 900	&2\arcsec$\times$6\farcm7	&103	&nucleus		\\
		&	 &2$\times$ 300	&2\arcsec$\times$6\farcm7	&103	&nucleus		\\
		&	 &6$\times$1200	&2\arcsec$\times$6\farcm7	 &13	&nucleus		\\
		&2s	 &1$\times$1800	&2\arcsec$\times$100\arcsec	 &13	&nucleus		\\
		&	 &1$\times$1800	&2\arcsec$\times$100\arcsec	 &13	&5\arcsec\ West of nucleus	\\
NGC 7213	&1s	 &6$\times$1800	&2\arcsec$\times$6\farcm7	&110	&along southern filament\\
\enddata
\tablecomments{
Col.(3): Total exposure time, in seconds.  Col.(4): Slit size (width $\times$ length), in arcseconds.  Col.(5): Position angle of slit, in degrees. Col. (6): Location of the slit. 
}
\end{deluxetable}


\begin{references}

\refpar
Adelberger, K. L., Steidel, C. C., Shapley, A. E., \& Pettini, M. 2003, \apj, 584, 45
\refpar
Antonucci, R. R. J., \& Miller, J. S. 1985, \apj, 297, 621
\refpar
Arribas, S., Mediavilla, E., \& Garcia-Lorenzo, B. 1996, \apj, 463, 509
\refpar
Bajtlik, S., Duncan, R. C., \& Ostriker, J. P. 1988, ApJ, 327, 570
\refpar
Baldwin, J., Wilson, A. S., \& Whittle, M. 1987, \apj, 319, 84
\refpar 
Bergeron, J., Boksenberg, A., Dennefeld, M., \& Tarenghi, M. 1983, \mnras, 
202, 125
\refpar
Bergeron, J., et al. 1994, \apj, 436, 33
\refpar
Beswick, R. J., Pedlar, A., Mundell, C. G., \& Gallimore, J. F. 2001, \mnras, 325, 151
\refpar 
Bland-Hawthorn, J. 1999, Nature, 400, 220
\refpar
Bland-Hawthorn, J., Freeman, K. C., \& Quinn, P. J. 1997a, \apj, 490, 143
\refpar
Bland-Hawthorn, J., Lumsden, S. L., Voit, G. M., Cecil, G. N., \& Weisheit, J. C. 1997b, Ap\&SS, 248, 177
\refpar
Bland-Hawthorn, J., \& Jones, D. H. 1998, PASA, 15, 44
\refpar
Bland-Hawthorn, J. \& Kedziora-Chudczer, L. 2003, PASA, in press (astro-ph/0305032)
\refpar
Bland-Hawthorn, J., Lumsden, S. L, Voit, G. M., Cecil, G. N., \& Weisheit, J. C. 1997, Ap\&SS, 248, 177
\refpar 
Bland-Hawthorn, J., Sokolowski, J., \& Cecil, G. 1991, \apj, 375, 78
\refpar
Bland-Hawthorn, J., Veilleux, S., Cecil, G. N., Putman, M. E., Gibson, B. K.,
 \& Maloney, P. R. 1998, \mnras, 299, 611
\refpar
Bland-Hawthorn, J., Wilson, A. S., \& Tully, R. B. 1991, \apj, 371, L19
\refpar 
Blitz, L., Spergel, D. N., Teuben, P. J., Hartmann, D., Burton, W. B. 1999, 
\apj, 514, 818
\refpar
B\"ohringer, H., Voges, W., Fabian, A. C., Edge, A. C., \& Neumann, D. M. 1993, MNRAS, 264, L25
\refpar
Burbidge, E. M., \& Burbidge, G. R. 1960, \apj, 132, 30
\refpar
Carral, P., Turner, J. L., \& Ho, P. T. P. 1990, \apj, 362, 434
\refpar 
Cecil, G., Bland, J., \& Tully, R. B. 1990, \apj, 355, 70
\refpar 
Cecil, G., Bland-Hawthorn, J., Veilleux, S., \& Filippenko, A. V. 2001, \apj, 
555, 338
\refpar 
Cecil, G., Bland-Hawthorn, J., \& Veilleux, S. 2002a, \apj, 576, 745
\refpar
Cecil, G, et al. 2000, \apj, 536, 675
\refpar
Cecil, G., et al. 2002b, \apj, 568, 627
\refpar
Cepa, J., et al. 1990, in Optical and IR Telescope Instrumentation and Detectors, eds. M. Iye and A. F. Moorwood, Proc. SPIE, 4008, 623
\refpar
Colina, L. 1992, \apj, 386, 59
\refpar
Crenshaw, D. M., \& Kraemer, S. B. 2000, \apj, 532, L101
\refpar
Dahlem, M., Ehle, M., Jansen, F., Heckman, T. M., Weaver, K. A.,
Strickland, D. K. 2003, \aap, 403, 547
\refpar
Dahlem, M., Lazendic, J. S., Haynes, R. F., Ehle, M., Lisenfeld, U. 2001,
\aap, 374, 42
\refpar
Dahlem, M., Petr, M. G., Lehnert, M. D., Heckman, T. M., Ehle, M. 1997,
\aap, 320, 731
%\refpar
%David, L. P., et al. 2001, ApJ, 557, 546
\refpar
de Vaucouleurs, G., de Vaucouleurs, A., Corwin, H. G. Jr., \& Buta,
R. J., Paturel, G., \& Fouqu\'e, P. 1991, Third Reference Catalogue of
Bright Galaxies (New York, Springer-Verlag)
\refpar 
Devine, D., \& Bally, J. 1999, \apj, 510, 197
\refpar
Dopita, M. A., Groves, B. A., Sutherland, R. S, Binette, L., \& Cecil, G.
2002, \apj, 572, 753
\refpar
Dopita, M. A., \& Sutherland, R. S. 1995, \apj, 455, 468
\refpar
Edmunds, M. G., Taylor, K., \& Turtle, A. J. 1988, \mnras, 234, 155
\refpar
Elmouttie, M., Haynes, R. F., Jones, K. L., Sadler, E. M., \& Ehle, M. 1998a, \mnras, 297, 1202
\refpar
Elmouttie, M., et al. 1998b, \mnras, 297, 49
\refpar
Fabian, A. C., et al. 2000, MNRAS, 318, L65
\refpar
Fabian, A. C. 2001, MNRAS, 321, L20 
\refpar
Falcke, H., Wilson, A. S., \& Simpson, C. 1998, \apj, 502, 199
\refpar
Ferguson, A., van der Hulst, T., \& van Gorkom, J. 2001, AAO Newsletter, 96, 4
\refpar
Ferland, G. J., \& Netzer, H. 1983, \apj, 264, 105 
\refpar
Filippenko, A. V., \& Halpern, J. P. 1984, \apj, 285, 458
\refpar 
Franx, M., \& Illingworth, G. 1990, \apj, 359, L41
\refpar
Freeman, K. C., et al. 1977, \aap, 55, 445
\refpar
Fried, J. W., \& Schulz, H. 1983, \aap, 118, 166
\refpar 
Fukugita, M., Hogan, C. J., \& Peebles, P. J. E. 1998, \apj, 503, 518
\refpar
Galletta, G., \& Recillas-Cruz, E. 1982, \aap, 112, 361
\refpar
Gallimore, J., et al. 1996, \apj, 464, 198
\refpar
Ganguly, R., Charlton, J. C., \& Eracleous, M. 2001, \apj, 556, L7
\refpar
Gerhard, O., Arnaboldi, M., Freeman, K. C., \& Okamura, S. 2002, \apj, 580, L121
\refpar
Glazebrook, K., \& Bland-Hawthorn, J. 2001, \pasp, 113, 197
\refpar 
Hameed, S., Blank, D. L., Young, L. M., \& Devereux, N. 2001, \apj, 546, L97
\refpar 
Heckman, T. M., Armus, L., \& Miley, G. K. 1987, \aj, 93, 276
\refpar 
-----. 1990, \apjs, 74, 833
\refpar
Heckman, T. M., \& Leitherer, C. 1997, \aj, 114, 69
\refpar
Heckman, T. M., et al. 2001, \apj, 554, 1021
\refpar
Heinz, S., Choi, Y.-Y., Reynolds, C. S., \& Begelman, M. C. 2002, \apj, 569, L79
\refpar
Helfer, T. T., \& Blitz, L. 1995, \apj, 450, 90 
\refpar
Hjelm, M., \& Lindblad, P. O. 1996, \aap, 305, 727
\refpar
Hummel, E., \& Saikia, D. J. 1991, \aap, 249, 43
\refpar
Hunter, D. A., Hawley, W. N., \& Gallagher, J. S. III 1993, \aj, 106, 1797
\refpar
Irwin, J. A., English, J., \& Sorathia, B. 1999, \aj, 117, 2102
\refpar
Iwasawa, K., \& Comastri, A. 1998, \mnras, 297, 1219
\refpar
Iwasawa, K., Wilson, A. S., Fabian, A. C., \& Young, A. J. 2003, \mnras, submitted
\refpar
Jenkins, A., et al. 1998, \apj, 499, 20
\refpar
J\"ors\"ater, S., Lindblad, P. O., \& Boksenberg, A. 1984, \aap, 140, 288
\refpar
Jones, D. H., Shopbell, P. L., \& Bland-Hawthorn, J. 2002, \mnras, 329, 759
\refpar
Kaneko, N., et al. 1992, AJ, 103, 422
\refpar
Keel, W. C. 1990, \aj, 100, 356
\refpar
Kim, D.-C., Sanders, D. B., Veilleux, S., Mazzarella, J. M., \& Soifer, B. T. 1995, \apjs, 98, 129
\refpar
Kinkhabwala, A., et al. 2002, \apj, 575, 732
\refpar
Kolaczyk, E. D., \& Dixon, D. D. 2000, \apj, 534, 490
\refpar
Komossa, St. 2001, \aap, 367, 801
\refpar
Komossa, St., Schulz, H., \& Greiner, J. 1998, \aap, 334, 110
\refpar
Komossa, St., et al. 2003, \apj, 582, L15
\refpar
Kraemer, S. B., Ruiz, J. R., \& Crenshaw, D. M. 1998, \apj, 508, 232
\refpar
Kristen, H., J\"orsater, S., Lindblad, P. O., \& Boksenberg, A. 1997, \aap, 328, 483
\refpar
Kukula, M. J., Pedlar, A., Baum, S. A., \& O'Dea, C. P. 1995, \mnras, 276, 1262
\refpar
Lanzetta, K. M., Bowen, D. V., Tytler, D., \& Webb, J. K. 1995, \apj, 442, 538
\refpar 
Larson, R. B., \& Dinerstein, H. L. 1975, \pasp, 87, 911
\refpar
Lehnert, M. D., \& Heckman, T. M. 1995, \apjs, 97, 89
\refpar
-----. 1996, \apj, 462, 651
\refpar 
Lehnert, M. D., Heckman, T. M., \& Weaver, K. A. 1999, \apj, 523, 575
\refpar
Lira, P., Ward, M. J., Zezas, A., \& Murray, S. S. 2002, \mnras, 333, 709
\refpar
Macchetto, F., Capetti, A., Sparks, W. B., Axon, D. J., \& Boksenberg, A. 1994, \apj, 435, L15
\refpar
Maiolino, R., Krabbe, A., Thatte, N., \& Genzel, R. 1998, \apj, 493, 650
\refpar
Marconi, A., Moorwood, A. F. M., Origlia, L., \& Oliva, E. 1994, Messenger, 78, 20
\refpar
Marlowe, A. T., Heckman, T. M., Wyse, R. F. G., \& Schommer, R. 1995, \apj, 438, 285
\refpar
McNamara, B. R., et al. 2000, ApJ, 534, L135
\refpar
Melnick, J., Moles, M., \& Terlevich, R. 1985, \aap, 149, L24
\refpar
Meurer, G. R., et al. 1999, BAAS, 194, 501
\refpar
Meurer, G. R., Freeman, K. C., Dopita, M. C., \& Cacciari, C. 1992, \aj, 103, 60
\refpar
Meurer, G. R., et al. 1995, \aj, 110, 2665
\refpar
Miller, J. S., \& Antonucci, R. R. J. 1983, \apj, 271, L7
\refpar
Miller, J. S., Goodrich, R. W., \& Mathews, W. G. 1991, \apj, 378, 47
\refpar
Miller, S. T., \& Veilleux, S. 2003a, ApJS, 148, 000
\refpar
-----. 2003b, ApJ, 592, 000
\refpar
Monier, E. M., Mathur, S., Wilkes, B., \& Elvis, M. 2001, \apj, 559, 675
\refpar
Morganti, R., Tsvetanov, Z. I., Gallimore, J., \& Allen, M. G. 1999, A\&AS, 137, 457
\refpar
Mulchaey, J. S., Wilson, A. S., \& Tsvetanov, Z. 1996, \apj, 467, 197
\refpar 
N\"orgaard-Nielsen, H. U., Hansen, L., J\"orgesen, H. E., \& Christensen, P. 
R. 1986, \aap, 169, 49
\refpar
Norman, C. A., Bowen, D. V., Heckman, T. M., Blades, C., \& Danly, L. 1996, \apj, 472, 73
\refpar
Petitjean, P., \& Durret, F. 1993, \aap, 277, 365
\refpar
Phillips, M. M. 1979, \apj, 227, L121
\refpar
Phillips, M. M., \& Malin, D. F. 1982, \mnras, 199, 205
\refpar
Phillips, M. M., Turtle, A. J., Edmunds, M. G., \& Pagel, B. E. J. 1983, \mnras, 203, 759
\refpar
Pogge, R. W. 1988a, \apj, 328, 519
\refpar
-----. 1988b, \apj, 332, 702
\refpar
Putman, M. E., Bland-Hawthorn, J., Veilleux, S., Gibson, B. K., Freeman, K. C., \& Maloney, P. R. 2003, \apj, submitted 
\refpar
Quilis, V., Bower, R. G., \& Balogh, M. L. 2001, MNRAS, 328, 1091
\refpar
Rauch, M. 1998, ARA\&A, 36, 267
\refpar
Reynolds, C. J. 1997, \mnras, 286, 513
\refpar
Ruiz, M., et al. 2000, \mnras, 316, 49
\refpar
Sahu, M. S., \& Blades, J. C. 1997, \apj, 484, L125
\refpar
Sambruna, R. M., et al. 2001a, \apj, 546, L9
\refpar
-----. 2001b, \apj, 546, L13
\refpar
Sandqvist, Aa., J\"ors\"ater, S., \& Lindblad, P. O. 1995, \aap, 295, 585
\refpar
Schulz, H., Komossa, S., Bergh\"ofer, T., \& Boer, B. 1998, \aap, 330, 823
\refpar
Scoville, N. Z., et al. 2000, \aj, 119, 991
\refpar
Scoville, N. Z., Matthews, K., Carico, D. P., \& Sanders, D. B. 1988, \apj, 327, L61
\refpar
Seyfert, C. K. 1943, \apj, 97, 28
\refpar
Shapley, A. E., Steidel, C. C., Pettini, M., \& Adelberger, K. L. 2003, \apj, 588, 65
\refpar
Shopbell, P. L., \& Bland-Hawthorn, J. 1998, \apj, 493, 129
\refpar 
Shopbell, P. L., Veilleux, S., \& Bland-Hawthorn, J. 1999, \apj, 524, L83
\refpar
Smith, D. A., \& Wilson, A. S. 2001, \apj, 557, 180
\refpar
Sofue, Y. 1997, PASJ, 49, 17
\refpar
Sofue, Y., et al. 1999, ApJ, 523, 136
\refpar
Soifer, B. T., Boehme, L., Neugebauer, G., \& Sanders, D. B. 1989, \aj, 98, 766
\refpar
Sokolowski, J., Bland-Hawthorn, J., \& Cecil, G. 1991, \apj, 375, 583
\refpar
Steidel, C. C., et al. 2002, ApJ, 576, 653
\refpar
Steidel, C. C., Pettini, M., Adelberger, K. L. 2001, ApJ, 546, 665
\refpar
Stevens, I. R., Read, A. M., \& Bravo-Guerrero, J. 2003, \mnras, preprint (astro-ph/0306334)
\refpar
Stone, J. L., Wilson, A. S., \& Ward, M. J. 1988, \apj, 330, 105
\refpar
Storchi-Bergmann, T., \& Bonatto, C. 1991, \mnras, 250, 138
\refpar
Storchi-Bergmann, T., Rodriguez-Ardila, A., Schmitt, H. R., Wilson, A. S., \& Baldwin, J. A. 1996, \apj, 472, 83
\refpar
Strickland, D. K., Heckman, T. M., Weaver, K. A., \& Dahlem, M. 2000, \aj, 120, 2965
\refpar
Tacconi, L., et al. 1999, \apj, 524, 732
\refpar
Tadhunter, C., Villar-Martin, M., Morganti, R., Bland-Hawthorn, J., \& Axon, D. 2000, \mnras, 314, 849
\refpar
Tezca, M., et al. 2000, \apj, 537, 178
\refpar
Tully, R. B. 1988, Nearby Galaxies Catalog, Cambridge University Press
\refpar
Thronson, H. A., et al. 1989, \apj, 343, 158
\refpar 
Vader, P. 1986, \apj, 305, 669
\refpar
Veilleux, S. 2002, in Extragalactic Gas at Low Redshift, 
ASP Conf. Proc. Vol. 254, eds. J. S. Mulchaey and J. Stocke, 
San Francisco: Astronomical Society of the Pacific, 313
\refpar
Veilleux, S., \& Bland-Hawthorn, J. 1997, \apj, 479, L105
\refpar
Veilleux, S., Bland-Hawthorn, J., Cecil, G., Tully, R. B., \& Miller, S. T. 1999, \apj, 520, 111
\refpar
Veilleux, S., Cecil, G., Bland-Hawthorn, J., Tully, R. B., Filippenko, A. V., 
\& Sarger, W. L. W. 1994, \apj, 433, 48
\refpar
Veilleux, S., Kim, D.-C., Sanders, D. B., Mazzarella, J. M., \& Soifer, B. T. 1995, \apjs, 98, 171
\refpar
Veilleux, S., \& Osterbrock, D. E. 1987, \apjs, 63, 295
\refpar
Veilleux, S., \& Rupke, D. S. 2002, \apj, 565, L63
\refpar
Vignati, P., et al. 1999, \aap, 349, L57
\refpar
Vollmer, B., \& Huchtmeier, W. 2003, \aap, in press (astro-ph/0303531) 
\refpar
Walker, M. 1968, \apj, 151, 71
\refpar
Weymann, R. J., Vogel, S. N., Veilleux, S., Epps, H. W. 2001, ApJ, 561, 559
\refpar
Wilson, A. S., et al. 2000, \aj, 120, 1325
\refpar 
Wilson, A. S., \& Tzvetanov 1994, \aj, 107, 1227
\refpar
Wilson, A. S., \& Ulvestad, J. 1987, \apj, 319, 105
\refpar
Yoshida, M., et al. 2002, \apj, 567, 118
\refpar
Young, A. J., Wilson, A. S., \& Shopbell, P. L. 2001, \apj, 556, 6

\end{references}
\end{document}